\documentclass[12pt]{article}

\usepackage{titlesec} 
\titleformat*{\section}{\large\bfseries}
\titleformat*{\subsection}{\large\bfseries}

\usepackage{hyperref} 
\usepackage[english]{babel}
\usepackage{mathtools, amssymb} 
\usepackage{mathtext} 
\usepackage{bm} 
\usepackage{amsthm} 
\usepackage{dsfont}
\usepackage{tikz,subcaption}
\usetikzlibrary{patterns, decorations.markings, arrows.meta}

\hypersetup{colorlinks=true, linktocpage=false, linkcolor=blue!80!black, citecolor=red!85!black, urlcolor=blue} 

\newtheorem{remark}{Remark}
	
\def\cbk{\color{black}}

\renewcommand{\imath}{i}
\renewcommand{\epsilon}{\varepsilon}
\renewcommand{\Re}{\operatorname{Re}}
\renewcommand{\Im}{\operatorname{Im}}
\renewcommand{\phi}{\varphi}

\DeclareMathOperator{\sign}{sign}
\DeclareMathOperator{\sh}{sh}
\DeclareMathOperator{\ctg}{ctg}
\DeclareMathOperator{\ch}{ch}

\tikzset{->-/.style={line width=0.25mm,decoration={
			markings,
			mark=at position 0.53 with {\arrow{Stealth[length=1.4mm,width=1.1mm]}}},postaction={decorate}}}

\numberwithin{equation}{section}

\usepackage{tocloft} 

\begin{document}
\thispagestyle{empty}
\vspace{2cm}

\begin{center}
{\bf \large
	Reflection operator and hypergeometry I:\\[6pt]
	$SL(2,\mathbb{R})$ spin chain}
\vspace{0.7cm}

{P. Antonenko$^{\dagger\diamond}$, N. Belousov$^{\dagger\circ}$, S. Derkachov$^{\dagger}$, S. Khoroshkin$^{\circ\ast}$}
\vspace{0.7cm}

{\small \it
	$^\dagger$Steklov Mathematical Institute, Fontanka 27, \\St.~Petersburg, 191023, Russia\vspace{0.3cm}\\
	$^\diamond$Leonhard Euler International Mathematical Institute,\\ Pesochnaya nab. 10, St.~Petersburg, 197022, Russia\vspace{0.3cm}\\
	$^\circ$National Research University Higher School of Economics,\\ Myasnitskaya 20, Moscow, 101000, Russia\vspace{0.3cm}\\
	$^\ast$Skolkovo Institute of Science and Technology,\\Skolkovo, 121205, Russia }
	
\vspace{1.6cm}

	{To the 90-th anniversary of L .D. Faddeev} 

\end{center}

\vspace{0.5cm}

\begin{abstract} \noindent
In this work we consider open $SL(2,\mathbb{R})$ spin chain, mainly the simplest case of one particle. Eigenfunctions of the model can be constructed using the so-called reflection operator. We~obtain several representations of this operator and show its relation to the hypergeometric function. Besides, we prove orthogonality and completeness~of \mbox{one-particle} eigenfunctions and connect them to the index hypergeometric transform. Finally, we briefly state the formula for the eigenfunctions in many-particle case.
\end{abstract}

\vspace{0.3cm}

\newpage
 \tableofcontents

 \newpage

\section{Introduction}

\subsection{Open spin chain}
In this paper we consider quantum integrable model called \textit{open $SL(2,\mathbb{R})$ spin chain}. To define this model we follow Sklyanin's approach~\cite{Skl}. The basic building block is the Lax matrix
\begin{align}\label{L}
	L(u) = \begin{pmatrix}
		u + S & S_- \\
		S_+ & u - S
	\end{pmatrix},
\end{align}
where $S, S_{\pm}$ are generators of the group $SL(2,\mathbb{R})$ satisfying commutation relations
\begin{equation}\label{Scomm}
	[S_+, S_-] = 2S, \qquad [S,S_\pm] = \pm S_\pm.
\end{equation}
For any representation of $SL(2,\mathbb{R})$ this Lax matrix solves the Yang-Baxter equation
\begin{align}\label{RLL}
R(u-v)\, \bigl(L(u)\otimes \bm{1}\bigr)\, \bigl( \bm{1} \otimes L(v)\bigr) =
\bigl( \bm{1} \otimes L(v)\bigr)\,\bigl(L(u)\otimes  \bm{1}\bigr)\, R(u-v)
\end{align}
with the Yang's $R$-matrix acting in $\mathbb{C}^2 \otimes \mathbb{C}^2$
\begin{align}\label{R}
	R(u)  =
	\begin{pmatrix}
		u + 1 & 0 & 0 & 0\\
		0 & u & 1 & 0\\
		0 & 1 & u & 0\\
		0 & 0 & 0 & u + 1
	\end{pmatrix}.
\end{align}
In what follows we consider representation such that the generators are realized by differential operators
\begin{equation}\label{S}
	S = z \partial_z + s, \qquad S_- = - \partial_z, \qquad S_+ = z^2 \partial_z + 2s z
\end{equation}
acting on functions $\psi(z)$ analytic in the upper half-plane~\mbox{$\Im z >0$.}
The generators are anti-hermitian
\begin{equation}
	S^\dagger = -S, \qquad S_-^\dagger = -S_-, \qquad S_+^\dagger = - S_+
\end{equation}
with respect to the scalar product
\begin{equation}\label{sc}
	\langle \chi| \psi\rangle = \int\limits \mathcal{D}z \; \overline{\chi(z)} \, \psi(z),
\end{equation}
where the integration is performed over upper half-plane $\Im z > 0$ and the measure is defined by the formula
\begin{align}\label{measure}
	\mathcal{D}z = \frac{2s - 1}{\pi} \, (2 \Im z)^{2s - 2} \; d\Re z \; d \Im z.
\end{align}
For the \textit{spin} parameter $s$ we assume
\begin{equation}
	s > \frac{1}{2}.
\end{equation}
The generators with integer or half-integer spin correspond to \textit{discrete series} of unitary irreducible representations of the group $SL(2,\mathbb{R})$ \cite[Ch. VII]{GGV}\cbk.

Lax matrix \eqref{L} contains two parameters~$u$~and~$s$.
In what follows we also use equivalent parametrization
\begin{align}
	L(u) \equiv L(u + s - 1, u - s),
\end{align}
where parameters $u_1 = u+s-1$,  $u_2 = u-s$ appear in a natural way in factorized expression
\begin{align}\nonumber
	L(u_1, u_2) &= \begin{pmatrix}
		1 & 0 \\
		z & 1
	\end{pmatrix}
	\begin{pmatrix}
		u_1 & - \partial_z \\
		0 & u_2
	\end{pmatrix}
	\begin{pmatrix}
		1 & 0 \\
		-z & 1
	\end{pmatrix} \\[6pt] \label{L2}
	&= \begin{pmatrix}
		u_1 + 1 + z \partial_z & - \partial_z \\[3pt]
		z^2 \partial_z + (u_1 - u_2 + 1) z & u_2 - z \partial_z
	\end{pmatrix}.
\end{align}
The second basic building block of the model is the $K$-matrix
\begin{align}\label{K}
	K(u) = \begin{pmatrix}
		\imath \alpha & u - \frac{1}{2} \\[6pt]
		-\beta^2 \bigl(u - \frac{1}{2} \bigr) & \imath \alpha
	\end{pmatrix},
\end{align}
which contains two additional parameters $\alpha, \beta$. The matrix $K(u)$ satisfies the reflection equation \cite{Ch, Skl, KS}
\begin{multline}\label{RKRK}
R(u-v)\, \bigl(K(u)\otimes \bm{1}\bigr)\,R(u+v-1)\, \bigl(\bm{1} \otimes K(v)\bigr) \\[6pt]
= \bigl(\bm{1} \otimes K(v)\bigr)\,R(u+v-1)\,\bigl(K(u)\otimes \bm{1}\bigr)\, R(u-v)
\end{multline}
with the same Yang's $R$-matrix \eqref{R}.

The open $SL(2,\mathbb{R})$ spin chain is a model of $n$ interacting particles with coordinates $z_j$. Denote by $L_j(u)$ Lax matrices corresponding to each particle
\begin{align}
	L_j(u) = \begin{pmatrix}
		u + z_j \partial_{z_j} + s & - \partial_{z_j} \\[6pt]
		z_j^2 \partial_{z_j} + 2s z_j & u - z_j \partial_{z_j} - s
	\end{pmatrix}.
\end{align}
Using the introduced objects we define the \textit{monodromy matrix}
\begin{align}\label{Tn}
	T(u) = L_n(u) \cdots L_1(u) K(u) L_1(u) \cdots L_n(u) = \begin{pmatrix}
		A(u) & B(u) \\
		C(u) & D(u)
	\end{pmatrix}.
\end{align}
Due to the equations for its building blocks \eqref{RLL}, \eqref{RKRK}, the monodromy matrix obeys the same reflection equation \cite[Proposition 2]{Skl}
\begin{multline}
R(u-v)\, \bigl(T(u)\otimes \bm{1}\bigr)\,R(u+v-1)\, \bigl(\bm{1} \otimes T(v)\bigr) \\[6pt]
= \bigl(\bm{1} \otimes T(v)\bigr)\,R(u+v-1)\,\bigl(T(u)\otimes \bm{1}\bigr)\, R(u-v).
\end{multline}
This identity is equivalent to the commutation relations between the elements of the monodromy matrix \eqref{Tn}. In particular, equality between $(14)$-elements in reflection equation gives commutativity
\begin{align}
	[B(u), B(v)] = 0.
\end{align}
Thus, the coefficients $H_k$ of the polynomial $B(u)$
\begin{align}\label{BH}
	B(u) =  \biggl(u - \frac{1}{2}\biggr)(u^{2n} + u^{2n - 2} H_1 + \ldots + H_n)
\end{align}
commute with each other
\begin{equation}
	[H_k, H_m] = 0
\end{equation}
and define the quantum integrable model.

Our main goal is the diagonalization of the operator $B(u)$. In this article we mostly focus on the case of one particle. In Section \ref{sec:many} we briefly state the answer for the general case and postpone its derivation to future paper.

\subsection{One-particle problem}

In this section we formulate our main results concerning one-particle problem. When $n = 1$ the operator $B(u)$ has the form
\begin{equation}
	B(u) =  \biggl(u - \frac{1}{2} \biggr) \bigl(u^2 - H^s\bigr)
\end{equation}
with the only nontrivial part being one-particle Hamiltonian
\begin{equation}\label{H}
	\begin{aligned}
		H^{s} &= S^2 + \beta^2 S_-^2 - 2\imath \alpha S_- \\[6pt]
		& =(z^2 + \beta^2) \partial_z^2 + (2s + 1) z \partial_z + 2 \imath \alpha \partial_z + s^2.
	\end{aligned}
\end{equation}
We look for the eigenfunctions of this operator
\begin{align}\label{HPsi}
	H^s \, \Psi_\lambda(z) = -\lambda^2 \, \Psi_\lambda(z)
\end{align}
analytic in the upper half-plane $\Im z >0$. The Hilbert space consists of such analytic functions square integrable with respect to the scalar product \eqref{sc}.

After rescaling and shifting
\begin{equation}\label{ch}
	w = \frac{1}{2} + \frac{\imath z}{2\beta}, \qquad z = \imath \beta (1 - 2w)
\end{equation}
the operator \eqref{H} transforms into
\begin{equation}\label{Hw}
	H^s = -w(1 - w)\partial_w^2 - \biggl[ \biggl( s + \frac{1}{2} + \frac{\alpha}{\beta} \biggr) - (2s + 1) w \biggr]\partial_w + s^2.
\end{equation}
Hence, the spectral problem \eqref{HPsi} can be rewritten as the hypergeometric equation
\begin{equation}\label{hyp}
	\Bigl( w(1 - w)\partial_w^2 + \bigl[ c - (a+b+1) w \bigr]\partial_w -ab  \Bigr)  \Psi_\lambda= 0
\end{equation}
with the parameters
\begin{equation}\label{abc}
	a = s + \imath \lambda, \qquad b = s - \imath \lambda, \qquad c = s + \frac{1}{2} + \frac{\alpha}{\beta}  .
\end{equation}
Therefore the eigenfunctions can be expressed in terms of hypergeometric function ${}_2 F_1(a, b,c;w)$, whose properties, of course, are very well-known \cite[\href{https://dlmf.nist.gov/15}{Ch. 15}]{DLMF}.

However, there is another approach to the spectral problem, which makes use of the underlying integrability of the model. The advantage of this approach is that it can be generalized to the case of many particles.

Before we describe it, let us make few remarks about the parameters of the model $\alpha, \beta$. In this paper we always assume
\begin{equation}\label{bg}
	\beta > 0, \qquad \frac{1}{2} + \frac{\alpha}{\beta} > 0.
\end{equation}
The reasons for these assumptions are as follows.
Spin operators are anti-hermitian $S^\dagger = -S$, $S_-^\dagger = -S_- $
with respect to the scalar product \eqref{sc}. Therefore the Hamiltonian \eqref{H} is formally self-adjoint
\begin{equation}
	(H^s)^\dagger = H^s
\end{equation}
under assumptions
\begin{equation}\label{ga}
	\alpha \in \mathbb{R}, \qquad \beta^2 \in \mathbb{R}.
\end{equation}
From this we have two options: $\beta \in \imath \mathbb{R}$ and $\beta \in \mathbb{R}$. The possible eigenfunctions in these cases are different due to analyticity requirement. Note that~$H^s$~\eqref{H} is invariant under reflection $\beta \to - \beta$. Thus, the general solution of the equation \eqref{hyp} can be written as the linear combination
\begin{equation}\label{psi}
	\begin{aligned}
		\Psi^{\mathrm{gen}}_\lambda(z) &=  A \; {}_2 F_1 \biggl( s + \imath \lambda, \, s - \imath \lambda, \,  s + \frac{1}{2} + \frac{\alpha}{\beta} ; \,  \frac{1}{2} + \frac{\imath z}{2\beta} \biggr)\\[6pt]
		& + B  \; {}_2 F_1 \biggl( s + \imath \lambda, \, s - \imath \lambda, \,  s + \frac{1}{2} - \frac{\alpha}{\beta} ; \,  \frac{1}{2} - \frac{\imath z}{2\beta} \biggr).
	\end{aligned}
\end{equation}
Since ${}_2F_1(a, b,c;w)$ has branch cut from $w=1$ to $\infty$, in the case $\beta \in \imath \mathbb{R}$ both linearly independent solutions in \eqref{psi} are analytic in the upper half-plane $\Im z > 0$. On the other hand, if $\beta \in \mathbb{R}$, then only one of them fits this requirement. For simplicity, in this article we consider only the second case and without loss of generality assume $\beta > 0$.

The second condition in \eqref{bg} is also imposed to simplify matters: in this case we exclude possibility of discrete spectra, see Remark \ref{rem2} in Section~\ref{sec:orth1}. In addition, it is convenient to denote
\begin{equation}
	g = \frac{1}{2} + \frac{\alpha}{\beta} > 0.
\end{equation}
In what follows we frequently use the parameter $g$ instead of $\alpha$.

So, under the above assumptions \eqref{bg} the eigenfunctions of $H^s$ analytic in the upper half-plane are given by the formula
\begin{align}\label{Psi}
	\Psi_{\lambda}(z) = {}_2F_1 \biggl( s + \imath \lambda, \, s - \imath \lambda, \,  s + g ; \,  \frac{1}{2} + \frac{\imath z}{2\beta} \biggr).
\end{align}
Notice that since $s + g > 1/2$ this solution is well-defined. Also observe that it is symmetric with respect to reflection $\lambda \to - \lambda$, so the corresponding eigenvalue $-\lambda^2$ \eqref{HPsi} is non-degenerate. 

Now let us describe an alternative way to construct the eigenfunctions. Imagine the operator $\mathcal{K}(s,x)$ acting on functions $\psi(z)$ and intertwining Hamiltonians \eqref{H} with different spins
\begin{align}\label{HKsx}
	H^s \, \mathcal{K}(s, x) = \mathcal{K}(s, x) \, H^x.
\end{align}
Then acting on $1$ from both sides we obtain
\begin{align}\label{HK}
	H^s \, \mathcal{K}(s,x) \cdot 1 = x^2 \, \mathcal{K}(s, x) \cdot 1,
\end{align}
and therefore $\mathcal{K}(s, x) \cdot 1$ is an eigenfunction. Such operator can be constructed by solving the equation
\begin{equation}\label{Kdef}
	\begin{aligned}
		&\mathcal{K}(s,x) \, L(u +x - 1, u - s) \, K(u) \, L(u  + s - 1, u - x) \\[6pt]
		&= L(u + s - 1, u - x) \, K(u) \, L(u + x - 1, u - s) \, \mathcal{K}(s,x)
	\end{aligned}
\end{equation}
with matrices defined in the previous section \eqref{L2}, \eqref{K}. This matrix relation is equivalent to three independent equations on $\mathcal{K}$-operator derived in Section~\ref{sec:refl}, see~\eqref{Krel}. One of them is precisely the intertwining property~\eqref{HKsx}. Since the above equation is very similar to the reflection equation on $K$-matrix \eqref{RKRK}, we call $\mathcal{K}(s,x)$ \textit{reflection operator}.

In this paper we derive three formulas for the reflection operator. The first one is obtained in Section \ref{sec:refl} by solving the defining equation \eqref{Kdef}
\begin{align}\label{Kf1}
	\mathcal{K}(s,x) = \frac{\Gamma \bigl( N + \frac{x - s}{2} + g \bigr)}{\Gamma \bigl( N + \frac{s - x}{2} + g \bigr)}, \qquad
	N = \frac{1}{2\imath \beta} \bigl[ (z^2 + \beta^2) \partial_z + (s + x)z \bigr].
\end{align}
From that in Section \ref{sec:K2} we derive the second formula
\begin{align}\label{Kf2}
	\mathcal{K}(s, x) = (2\imath \beta)^{s - x} \,
	(z + \imath \beta)^{g-s}\,
	\frac{ \Gamma \bigl( (z - \imath \beta) \partial_z + x + g \bigr) }
	{ \Gamma \bigl( (z - \imath \beta) \partial_z + s + g \bigr) } \,
	(z + \imath \beta)^{x - g}.
\end{align}
The explicit action of such operators on $\psi(z)$ analytic in the upper half-plane can be written using the standard beta function integral~\cite[\href{http://dlmf.nist.gov/5.12.E1}{(5.12.1)}]{DLMF}. For the second one~\eqref{Kf2} we have
\begin{align}\label{Kf2-2}
	\begin{aligned}
		\bigl[\mathcal{K}(s,x) \, \psi \bigr](z) &= \frac{(2\imath \beta)^{s - x}}{\Gamma(s - x)} \, (z + \imath \beta)^{g - s} \, \int_0^1 dt \; (1 - t)^{s - x - 1} \; t^{g+x - 1} \\[6pt]
		& \hspace{2cm} \times  \bigl( t(z - \imath \beta) + 2\imath \beta\bigr)^{x - g} \; \psi\bigl(t(z-\imath \beta) + \imath \beta\bigr),
	\end{aligned}
\end{align}
and for the first one see \eqref{K11}. Using the last formula in Section \ref{sec:hyper} we prove that reflection operator produces the same eigenfunctions $\Psi_\lambda(z)$ \eqref{Psi}
\begin{align}
	\mathcal{K}(s, \imath \lambda) \cdot 1 = C(s, \imath \lambda) \; {}_2 F_1 \biggl(s + \imath \lambda, \, s - \imath \lambda, \, s + g ; \, \frac{1}{2} + \frac{\imath z}{2\beta} \biggr)
\end{align}
with the normalization
\begin{equation}
C(s, \imath \lambda) = \frac{\Gamma ( g+\imath \lambda)}
{\Gamma ( g+s )}.
\end{equation}
Finally, in Section \ref{sec:K3} we derive the third expression for the reflection operator with the integrals over upper half-plane
\begin{equation}\label{Kf3}
	\begin{aligned}
		&\bigl[\mathcal{K}(s, x) \, \psi \bigr] (z) = e^{2 \pi \imath s}\,(2\imath \beta)^{s - x}\,
		\frac{\Gamma(g+x) \, \Gamma(3s-g)}{\Gamma^2(2s)}\, (z + \imath \beta)^{g - s}  \\[6pt]
		& \times \int \mathcal{D} w \, \mathcal{D} v \; (z-\bar{v})^{-g-x} \, (i\beta-\bar{v})^{x-s} \, (v-\bar{w})^{g - 3s} \, (w + \imath \beta)^{x - g} \, \psi(w).
	\end{aligned}
\end{equation}
In Appendix \ref{Diagrammar} we describe how such integrals can be represented by diagrams. In particular, we obtain diagrammatic representation for the reflection operator \eqref{Kf3} and the eigenfunctions \eqref{Psi}, see Figures~\ref{fig:K-operator} and~\ref{fig:2F1}. In this approach various properties of the eigenfunctions reduce to simple transformations of corresponding diagrams. This technique was crucial in the study of other $SL(2,\mathbb{R})$ spin chains \cite{DKM1, DKM2}.

The formula for the reflection operator equivalent to \eqref{Kf1} has already appeared in the work \cite[eq. (3.23), (3.25)]{FGK}, which we discovered upon completing our paper. Since the derivation in \cite{FGK} is quite different, we leave our presentation as it is, and clarify the connection with the paper~\cite{FGK} in Appendix \ref{App:FGK}. We also remark that in \cite{FGK} this reflection operator was used in a different way, namely, to construct local Hamiltonian with interaction at the boundaries for a spin chain with many particles.

To our knowledge the two other formulas for the reflection operator~\eqref{Kf2-2}, \eqref{Kf3} are new. The close relative of the second formula~\eqref{Kf2-2} plays the key role in the case of open $SL(2,\mathbb{C})$ spin chain \cite[Section 3]{ABDV}.

In Section \ref{sec:orth} we show that the eigenfunctions $\Psi_{\lambda}(z)$ with $\lambda \in \mathbb{R}$ are orthogonal with respect to the scalar product \eqref{sc}
\begin{align}
	\langle \Psi_\rho | \Psi_\lambda \rangle = \mu^{-1}(\lambda) \; \frac{\delta(\lambda - \rho) + \delta(\lambda + \rho)}{2},
\end{align}
where the normalization coefficient
\begin{align}
	\mu(\lambda) = \frac{1}{4 \pi \, (2\beta)^{2s} \, \Gamma(2s)} \; \biggl| \frac{\Gamma^2(s + \imath \lambda) \, \Gamma ( g+\imath \lambda ) }{\Gamma ( s + g ) \, \Gamma(2\imath \lambda)}\biggr|^2.
\end{align}
We give two proofs of the orthogonality: from the asymptotics of eigenfunctions and in the language of diagrams.

In Section \ref{sec:compl} using Barnes representation of hypergeometric function we prove the completeness relation
\begin{equation}
	\int_\mathbb{R} d\lambda \; \mu(\lambda)  \; \Psi_\lambda(z) \, \overline{\Psi_\lambda(w)}= \frac{e^{\imath \pi s}}{(z - \bar{w})^{2s}}.
\end{equation}
Here from the right we have the kernel of the identity operator on our Hilbert space
\begin{equation}\label{Rep}
	\psi(z) = \int \mathcal{D} z \, \frac{e^{\imath \pi s}}{(z - \bar{w})^{2s}} \, \psi(w),
\end{equation}
the so-called \textit{reproducing kernel} \cite[(A.7)]{DKM1}.

Orthogonality and completeness are equivalent to the unitarity of the transform
\begin{equation}
	[ T \psi ] (\lambda) = \int \mathcal{D}z \; \overline{\Psi_\lambda(z)} \, \psi(z)
\end{equation}
that maps the initial Hilbert space to the space of functions of $\lambda \in \mathbb{R}$ square integrable with respect to the measure $\mu(\lambda) \, d\lambda$ and invariant under reflection~$\lambda \to -\lambda$. Interestingly enough, the transform $T$ has almost the same kernel as the well-known \textit{index hypergeometric (or Jacobi) transform} \cite{N}
\begin{align}
	[ J \psi ] (\lambda) = \int_0^\infty dy \,\, y^{s + g - 1}  \, (1 + y)^{s - g} 
	\; \overline{\Psi_\lambda \bigl(\imath \beta(1 + 2y)\bigr)} \; \psi(y),
\end{align}
which is also unitary, but acts on the different space of functions. In Section~\ref{sec:ind} we discuss the relation between these transforms.

\subsection{Many-particle problem}\label{sec:many}

In this section we state the result for the eigenfunctions in the case~of~$n$~particles
\begin{equation}\label{BPsi-n}
	B(u) \, \Psi_{\lambda_1, \ldots, \, \lambda_n} = \biggl(u - \frac{1}{2} \biggr)\,\prod_{j = 1}^n (u^2 + \lambda_j^2) \; \Psi_{\lambda_1, \ldots, \, \lambda_n}.
\end{equation}
For this, apart from reflection operator, we need $\mathcal{R}$-operator acting on functions of two variables $z_i$ and $z_j$
\begin{align}
	\mathcal{R}_{ij}(s, x) = \frac{\Gamma\bigl( (z_i - z_j)\partial_{z_i} + s + x \bigr)}
{\Gamma\bigl( (z_i - z_j )\partial_{z_i} + 2s \bigr)}.
\end{align}
The explicit formula for its action can be obtained using beta function integral. We note in passing that it interchanges the arguments in the following Lax matrices
\begin{align}
	\begin{aligned}
			& \mathcal{R}_{12}(s, x) \, L_1(u + s - 1, u - s) \, L_2(u + s - 1, u - x) \\[6pt]
			& = L_1(u + s - 1, u - x) \, L_2(u + s - 1, u - s) \, \mathcal{R}_{12}(s, x).
	\end{aligned}
\end{align}
Using $\mathcal{R}$- and $\mathcal{K}$-operators we introduce an infinite-dimensional analog of the monodromy matrix \eqref{Tn}
\begin{align}
	\Lambda_n(x) = \mathcal{R}_{n \, n - 1}(x) \cdots \mathcal{R}_{21}(x) \, \mathcal{K}_1(x) \, \mathcal{R}_{12}(x) \cdots \mathcal{R}_{n - 1 \, n}(x),
\end{align}
where the operator $\mathcal{K}_1(x)$ acts on functions of $z_1$ and in all operators we suppressed the dependence on spin $s$. Then eigenfunctions \eqref{BPsi-n} are given by the formula
\begin{align}
	\Psi_{\lambda_1, \ldots, \, \lambda_n}(z_1, \ldots, z_n) =\Lambda_n( \imath \lambda_n) \,\Lambda_{n - 1}( \imath \lambda_{n - 1}) \cdots \Lambda_1(\imath \lambda_1) \cdot 1.
\end{align}
The proof will be given elsewhere. Note that for the simplest $K$-matrix $K(u) = \bm{1}$ such construction of eigenfunctions is given in \cite{DKM2}.

\section{Reflection operator}\label{sec:K}
\subsection{Solving reflection equation}\label{sec:refl}
The operator $\mathcal{K}(s,x)$ acts on functions $\psi(z)$ and satisfies reflection equation
\begin{align}\label{Keq}
	\begin{aligned}
		&\mathcal{K}(s, x) \, L(u +x - 1, u - s) \, K(u) \, L(u  + s - 1, u - x) \\[6pt]
		&= L(u + s - 1, u - x) \, K(u) \, L(u + x - 1, u - s) \, \mathcal{K}(s,x),
	\end{aligned}
\end{align}
where $2\times 2$ matrices $L(u_1, u_2)$ and $K(u)$ are defined in \eqref{L2} and \eqref{K}. This matrix equation is equivalent to four equations between matrix elements. Each matrix element in turn is polynomial in $u$. Since $\mathcal{K}(s,x)$ doesn't depend on $u$, for each matrix element we obtain several equations corresponding to different degrees of $u$.

To find these equations, first, observe that Lax matrices can be written~as
\begin{align}
	& L(u +x - 1, u - s) =
		\begin{pmatrix}
			u + \frac{x - s}{2} + J & J_- \\[8pt]
			J_+ & u + \frac{x - s}{2} - J
		\end{pmatrix}, \\[8pt]
	& L(u +s- 1, u - x) =
	\begin{pmatrix}
		u - \frac{x - s}{2} + J & J_- \\[8pt]
		J_+ & u - \frac{x - s}{2} - J
	\end{pmatrix},
\end{align}
where $J, J_\pm$ are $SL(2, \mathbb{R})$ generators of spin $(s + x)/2$
\begin{equation}\label{J}
	J = z \partial_z + \frac{s + x}{2}, \qquad J_- = - \partial_z, \qquad J_+ = z^2 \partial_z + (s + x)z.
\end{equation}
They satisfy standard commutation relations
\begin{align}\label{Jcomm}
	[J_+, J_-] = 2J, \qquad [J, J_\pm] = \pm J_\pm.
\end{align}
Note also that the Casimir element
\begin{equation}\label{casimir}
	C = 2J^2 + J_+ J_- + J_- J_+ = \frac{(s + x)(s + x - 2)}{2}
\end{equation}
is symmetric with respect to $s, x$.

On the other hand, products of matrices from opposite sides of equation~\eqref{Keq} differ only by interchange of parameters $s \leftrightarrows x$. Using commutation relations \eqref{Jcomm} and formula for the Casimir element \eqref{casimir} one can check that all matrix elements in \eqref{Keq} are divisible by $(u - 1/2)$ up to some constants symmetric with respect to $s, x$.

Dividing each matrix element from \eqref{Keq} by $(u - 1/2)$ we compare coefficients behind remaining degrees of $u$. Their equality reduces to the following three relations
\begin{align}\label{Krel}
	\mathcal{K} \, N = N \, \mathcal{K}, \qquad \mathcal{K} \, H^x = H^s \, \mathcal{K}, \qquad \mathcal{K} \, I^{x, s} = I^{s, x} \, \mathcal{K}
\end{align}
with the operators
\begin{align}\label{N}
	& N = \frac{1}{2 \imath \beta} (J_+ - \beta^2 J_-), \\[6pt] \label{H2}
	& H^{s} = \Bigl( J + \frac{s - x}{2} \, \Bigr)^2 + \beta^2 J_-^2 - 2 \imath \alpha J_-, \\[6pt] \label{Isx}
	& I^{s, x} = J J_+ - \beta^2 J_- J + \frac{s - x}{2}(J_+ + \beta^2 J_-) + 2\imath \alpha J.
\end{align}
Note that the first operator is symmetric with respect to $s, x$
\begin{align}
	N = \frac{1}{2\imath \beta} \bigl[ (z^2 + \beta^2) \partial_z + (s + x)z \bigr].
\end{align}
The second operator $H^s$ doesn't depend on $x$ and coincides with one-particle Hamiltonian \eqref{H}.

The reflection operator commutes with $N$ \eqref{Krel}. Hence, we can search for it in the form
\begin{equation}
	\mathcal{K} = f(N).
\end{equation}
Let us derive equation on the function $f$ from the two remaining relations in~\eqref{Krel}. Denote
\begin{equation}
	N_\pm = - J \pm \frac{1}{2\imath \beta} (J_+ + \beta^2 J_-).
\end{equation}
The operators $N, N_\pm$ are linear combinations of generators $J, J_\pm$ and satisfy the same commutation relations \eqref{Jcomm}
\begin{align}
	[N_+, N_-] = 2N, \qquad [N, N_\pm] = \pm N_\pm.
\end{align}
From the second relation we deduce
\begin{equation}\label{fpm}
	f(N) \, N_\pm = N_\pm \, f(N \pm 1).
\end{equation}
Now we express old generators in terms of new ones
\begin{align}
	& J_+ = \frac{\imath \beta}{2  }(N_+ - N_- + 2N), \\[6pt]
	& J_- = \frac{\imath}{2  \beta} (N_+ - N_- - 2N), \qquad J = -\frac{1}{2}(N_+ + N_-)
\end{align}
and using these relations rewrite operators $H^s$ \eqref{H2} and $I^{s,x}$ \eqref{Isx} in terms of new generators
\begin{align}\nonumber
	& H^s = \biggl( N + \frac{x - s - 1}{2} + \frac{\alpha}{\beta} \biggr) N_+ + \biggl( - N + \frac{x - s - 1}{2} - \frac{\alpha}{\beta} \biggr) N_- \\[6pt]
	& \hspace{0.55cm} - 2N^2 - \frac{2\alpha}{\beta}N + \frac{s^2 + x^2 - s - x}{2},\\[10pt] \nonumber
	& \frac{1}{\imath \beta} \, I^{s, x}  =  - \biggl( N + \frac{x - s - 1}{2} + \frac{\alpha}{\beta} \biggr) N_+ + \biggl( - N + \frac{x - s - 1}{2} - \frac{\alpha}{\beta} \biggr) N_- \\[6pt]
	&\hspace{1.2cm} + N.
\end{align}
With the help of these formulas and formula \eqref{fpm} we reduce two relations
\begin{equation}
	f(N) \, H^x = H^s \, f(N), \qquad f(N) \, I^{x,s} = I^{s,x} \, f(N)
\end{equation}
to the single difference equation
\begin{align}
	\frac{f(N + 1)}{f(N)} = \frac{N + \frac{x - s + 1}{2} + \frac{\alpha}{\beta} }{ N + \frac{s - x + 1}{2} + \frac{\alpha}{\beta}}.
\end{align}
Its solution can be written in terms of gamma functions
\begin{align}\label{f}
	\mathcal{K}(s,x) = f(N) = \frac{\Gamma \bigl( N + \frac{x - s}{2} + g \bigr)}
{\Gamma \bigl( N + \frac{s - x}{2} + g \bigr)}.
\end{align}
where $g = 1/2 + \alpha/\beta$.

Of course, the expression \eqref{f} multiplied by any periodic function (with unit period) also solves the above equation. In particular, we can take function $f(N)$ with reflected parameter $\beta \to - \beta$ since the initial equation \eqref{Keq} depends on $\beta^2$. We chose specifically the solution \eqref{f}, because in the case
\begin{equation}\label{b}
	\beta > 0
\end{equation}
it gives us analytic in the upper half-plane eigenfunctions of one-particle Hamiltonian $H^s$ \eqref{H}, see Section \ref{sec:hyper}.

\subsection{Beta integral representations}\label{sec:K2}
In this section we obtain explicit formula for the action of the operator $\mathcal{K}(s,x)$ found in the previous section
\begin{align}\label{Ksx}
	\mathcal{K}(s,x) = \frac{\Gamma \bigl( N + \frac{x - s}{2} + g \bigr)}
	{\Gamma \bigl( N + \frac{s - x}{2} + g \bigr)}, \qquad
	N = \frac{1}{2\imath \beta} \bigl[ (z^2 + \beta^2) \partial_z + (s + x)z \bigr].
\end{align}
Moreover, to relate this operator to the hypergeometric function we derive another formula for it
\begin{align}\label{K2}
\mathcal{K}(s, x) = (2\imath \beta)^{s - x} \,
(z + \imath \beta)^{g-s}
\frac{ \Gamma \bigl( (z - \imath \beta) \partial_z + x + g \bigr) }
{ \Gamma \bigl( (z - \imath \beta) \partial_z + s + g \bigr) } \,
(z + \imath \beta)^{x - g}.
\end{align}
For both tasks we use the Euler's beta integral \cite[\href{http://dlmf.nist.gov/5.12.E1}{(5.12.1)}]{DLMF}.

First, notice that
\begin{align}
	\begin{aligned}
		N  &= \frac{1}{2\imath \beta} \bigl[ (z^2 + \beta^2) \partial_z + (s + x)z \bigr]  \\[6pt]
		& = (z + \imath \beta)^{-s - x} \; \biggl( \frac{1}{2\imath \beta}  (z^2 + \beta^2) \partial_z + \frac{s + x}{2} \biggr) \; (z + \imath \beta)^{s + x}.
	\end{aligned}
\end{align}
As a result, the expression \eqref{Ksx} is equivalent to
\begin{align}
	\begin{aligned}
		\mathcal{K}&=  (z + \imath \beta)^{-s - x} \; \frac{\Gamma \bigl( \frac{1}{2\imath \beta} (z^2 + \beta^2) \partial_z + x + g \bigr)}
{\Gamma \bigl(  \frac{1}{2\imath \beta} (z^2 + \beta^2) \partial_z + s + g \bigr)} \; (z + \imath \beta)^{s + x}.
	\end{aligned}
\end{align}
The second step is to use Euler's beta integral for the ratio of two gamma functions
\begin{align}\label{Kint}
	\begin{aligned}
		\mathcal{K} =  \frac{1}{\Gamma(s - x)} \, (z + \imath \beta)^{-s - x} \, \int_0^1 & dt \; (1 - t)^{s - x -1} \, t^{g + x - 1} \\[6pt]
		& \times   t^{ \frac{1}{2\imath \beta} (z^2 + \beta^2) \partial_z } \; (z + \imath \beta)^{s + x}.
	\end{aligned}
\end{align}
Under conformal map
\begin{align}
	\zeta = \frac{z - \imath \beta}{z + \imath \beta}, \qquad z = \imath \beta \, \frac{1 + \zeta}{1 - \zeta}
\end{align}
the operator in the integrand \eqref{Kint} drastically simplifies
\begin{align}
	\frac{1}{2\imath \beta} (z^2 + \beta^2) \partial_z = \zeta \partial_\zeta.
\end{align}
Hence, acting on arbitrary function $\psi(z)$ we obtain
\begin{align}
		t^{ \frac{1}{2\imath \beta} (z^2 + \beta^2) \partial_z } \, \psi(z) &= t^{\zeta \partial_\zeta} \; \psi \biggl( \imath \beta \, \frac{1 + \zeta}{1 - \zeta} \biggr) = \psi \biggl( \imath \beta \, \frac{ z + \imath \beta + t(z - \imath \beta) }{ z + \imath \beta - t(z - \imath \beta) } \biggr).
\end{align}
Note that for $\Im z \geq 0$, $t \in [0,1]$ and $\beta > 0$ we have
\begin{align}\label{Im}
	\Im  \biggl( \imath \beta \, \frac{ z + \imath \beta + t(z - \imath \beta) }{ z + \imath \beta - t(z - \imath \beta) }  \biggr) \geq 0,
\end{align}
and equality holds iff $\Im z = 0$, $t = 1$. Thus, for the function $\psi(z)$ analytic in the upper half-plane we stay in the region of analyticity. Acting on such function with operator \eqref{Kint} we arrive at
\begin{align}\label{K11}
	\begin{aligned}
		\mathcal{K} \, \psi(z) &=  \frac{(2\imath \beta)^{s + x}}{\Gamma(s - x)}  \, \int_0^1  dt \; (1 - t)^{s - x -1} \, t^{g+x - 1} \\[6pt]
		& \times   \bigl(z + \imath \beta - t(z - \imath \beta) \bigr)^{-s - x} \; \psi \biggl( \imath \beta \, \frac{ z + \imath \beta + t(z - \imath \beta) }{ z + \imath \beta - t(z - \imath \beta) }  \biggr).
	\end{aligned}
\end{align}
This formula represents the explicit action of the operator \eqref{Ksx}.

To derive the second formula \eqref{K2} we change integration variable $t \to u$
\begin{align}\label{ut}
	u = \frac{2 \imath \beta \, t}{z + \imath \beta - t(z - \imath \beta)}.
\end{align}
After this change we obtain
\begin{align}
	\begin{aligned}
		\mathcal{K} \, \psi(z) = \frac{(2\imath \beta)^{s - x}}{\Gamma(s - x)} \, &(z + \imath \beta)^{g - s} \, \int_{C} du \; (1 - u)^{s - x - 1} \; u^{g+x - 1} \\[6pt]
		& \times \bigl( u(z - \imath \beta) + 2\imath \beta \bigr)^{x - g} \; \psi \bigl( u(z - \imath \beta) + \imath \beta \bigr)
	\end{aligned}
\end{align}
with some contour ${C}$ from $u = 0$ to $u = 1$. Notice that the second line can be equivalently written as
\begin{multline}\label{udz}
	\quad \bigl( u(z - \imath \beta) + 2\imath \beta \bigr)^{x - g} \; \psi \bigl( u(z - \imath \beta) + \imath \beta \bigr)\\[6pt]
	= u^{(z - \imath \beta)\partial_z} \; (z + \imath \beta)^{x - g} \; \psi(z). \quad 
\end{multline}
Let us argue that the contour ${C}$ can be deformed to the interval $[0,1]$ without touching any singular points of the integrand. For $\Im z \geq 0$, $t \in [0,1]$ and $\beta > 0$ the new variable $u$ \eqref{ut} has properties
\begin{align}
	\Re u \in [0, 1], \qquad \sign ( \Im u) = \sign ( \Re z ).
\end{align}
Therefore
\begin{align}
	\Im \bigl( u(z - \imath \beta) + \imath \beta \bigr) = \beta ( 1 - \Re u) + \Im z \Re u + \Re z \Im u \geq 0,
\end{align}
and by decreasing $| \Im u \, | \searrow 0$ we stay in the analyticity domain of the function
\begin{align}
	\bigl( u(z - \imath \beta) + 2\imath \beta \bigr)^{x - g} \; \psi \bigl( u(z - \imath \beta) + \imath \beta \bigr).
\end{align}
So, deforming contour we finally arrive at the formula
\begin{align}\label{Kpsi}
	\begin{aligned}
		\mathcal{K} \, \psi(z) = \frac{(2\imath \beta)^{s - x}}{\Gamma(s - x)} \,  (z + \imath \beta)^{g - s} \, \int_0^1 & du \; (1 - u)^{s - x - 1} \; u^{g+x - 1} \\[6pt]
		& \times u^{(z - \imath \beta)\partial_z} \; (z + \imath \beta)^{x - g} \; \psi(z),
	\end{aligned}
\end{align}
which is equivalent to \eqref{K2} .

\subsection{Diagrammatic representation}\label{sec:K3}

In Appendix \ref{Diagrammar} we describe useful diagrammatic approach that allows to reduce calculations of integrals to simple transformations of corresponding diagrams. All diagrams represent the integrals over upper half-plane with the measure \eqref{measure}.

 \begin{figure}[t]
	\centering
	\begin{tikzpicture}[thick, line cap = round]
		\def\l{2}
		\def\r{1.5pt}
		\draw[->-] (0,0) node[left] {$z$} to node[midway,above=2pt]{\footnotesize $g+x$} (\l,0);
		\draw[fill = black] (\l,0) circle (\r);
		\draw[->-] (\l,0) to node[midway,above=2pt]{\footnotesize $3s - g$} (2*\l,0) node[right]{$w$};
		\draw[->-] (\l, \l) node[above]{$\imath \beta$} to node[pos = 0.35, left=2pt]{\footnotesize $s - x$} (\l, 0);
		\draw[->-] (0,0) to node[pos = 0.65, left = 6pt]{\footnotesize $s - g$} (\l, - \l);
		\draw[->-] (2*\l, 0) to node[pos = 0.65, right = 6pt]{\footnotesize $g-x$} (\l, -\l) node[below = 1pt]{$-\imath \beta$};
	\end{tikzpicture}
	\caption{Kernel of the reflection operator} \label{fig:K-operator}
\end{figure}
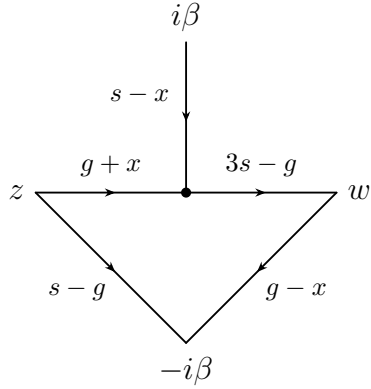

In this section we obtain diagrammatic representation for the kernel of reflection operator, see Figure \ref{fig:K-operator}. This diagram corresponds to the following formula
\begin{equation}\label{Kf3-2}
	\begin{aligned}
		&\bigl[\mathcal{K}(s, x) \, \psi \bigr] (z) =  e^{2\pi i s}  \,(2\imath \beta)^{s - x}\,
		\frac{\Gamma(g+x) \Gamma(3s-g)}{\Gamma^2(2s)} \, (z + \imath \beta)^{g - s} \\[8pt]
		&\quad \times\int \mathcal{D} w \, \mathcal{D} v \; \frac{\psi(w)}{  (z-\bar{v})^{g+x} \, (i\beta-\bar{v})^{s-x} \, (v-\bar{w})^{3s - g} \, (w + \imath \beta)^{g - x} }
	\end{aligned}
\end{equation}
up to coefficient behind the integral.

To derive \eqref{Kf3-2} we start from the formula \eqref{Kpsi}. The reflection operator is factorized in a product of three operators, so that its integral kernel admits similar factorization
\begin{align}\label{Kintw}
\mathcal{K} \, \psi(z) = \int \mathcal{D} w\,
(z + \imath \beta)^{g - s}\,
\mathcal{K}(z,\bar{w})\, (w + \imath \beta)^{x - g}\, \psi(w).
\end{align}
To obtain explicit form of the function $\mathcal{K}(z, \bar{w})$ we use reproducing kernel~\eqref{Rep} and rewrite the integral \eqref{Kpsi} in the form
\begin{align}
	\begin{aligned}
		\mathcal{K} \, \psi(z) &= \frac{(2\imath \beta)^{s - x}}{\Gamma(s - x)}
		\, (z + \imath \beta)^{g - s}\,
		\int_0^1 du \; (1 - u)^{s - x - 1} \; u^{g+x-1} \\[6pt]
		&\times u^{(z - \imath \beta)\partial_z} \;
		\int \mathcal{D} w\,
		\frac{e^{i\pi s}}{(z-\bar{w})^{2s}}\,
		(w + \imath \beta)^{x - g}\,\psi(w).
	\end{aligned}
\end{align}
In the last line we can use the formula
\begin{align}
u^{(z - \imath \beta)\partial_z}\,
\frac{1}{(z-\bar{w})^{2s}} =
\frac{1}{(u(z-\imath \beta)+\imath \beta-\bar{w})^{2s}}.
\end{align}
Then the function $\mathcal{K}(z, \bar{w})$ defined by \eqref{Kintw} admits the following expression
\begin{align}
\mathcal{K}(z,\bar{w}) =  e^{i\pi s}  \, \frac{(2\imath \beta)^{s - x}}{\Gamma(s - x)}\,
\int_0^1 du\,\frac{(1 - u)^{s - x - 1}\,u^{g+x-1}}
{(u(z-\imath \beta)+\imath \beta-\bar{w})^{2s}} .
\end{align}
The last step is to use the identity~\eqref{w-abc} proven in Appendix \ref{Diagrammar}
\begin{align}
	\begin{aligned}
		&\int \mathcal{D} v \; \frac{1}{(z-\bar{v})^b(i\beta-\bar{v})^{c-b}(v-\bar{w})^{2s + a-c}} \\[8pt]
		& \quad = e^{-i\pi s}\,\frac{\Gamma(a)\Gamma(2s)}{\Gamma(b)\Gamma(c-b) \Gamma(2s + a - c)} \,  \int_0^1 du \; \frac{(1-u)^{c-b-1} \, u^{b-1}}{ \bigl( u(z - \imath \beta) + \imath \beta  - \bar{w}\bigr)^{a}}
	\end{aligned}
\end{align}
and arrive at the integral over upper half-plane
\begin{align}
	\begin{aligned}
		\mathcal{K}(z,\bar{w}) &=  e^{2  i \pi s}  \, (2\imath \beta)^{s - x}\,
		\frac{\Gamma(g+x) \Gamma(3s-g)}{\Gamma^2(2s)} \\[6pt]
		& \times \int \mathcal{D} v \; \frac{1}{(z-\bar{v})^{g + x} \, (i\beta-\bar{v})^{s - x} \, (v-\bar{w})^{3s - g}}.
	\end{aligned}
\end{align}
Together with the formula \eqref{Kintw} it gives the result \eqref{Kf3-2}.

\section{Relation to hypergeometric function}\label{sec:hyper}

In this section we prove that action of the reflection operator on $1$ gives the~hypergeometric function
\begin{equation}\label{K2F1}
	\mathcal{K}(s, x) \cdot 1 = C(s,x) \; {}_2 F_1 \biggl(s + x, s - x, s + g ; \, \frac{1}{2} + \frac{\imath z}{2\beta} \biggr)
\end{equation}
with normalization
\begin{equation}
	C(s, x) =\frac{\Gamma ( g + x)}{\Gamma ( g + s)} .
\end{equation}
Starting from the representation \eqref{K2}, or equivalently formula \eqref{Kpsi}, we obtain
\begin{align}
	\begin{aligned}
		\mathcal{K}(s,x) \cdot 1 &= \frac{(2\imath \beta)^{s - g}}{\Gamma(s - x)} \, (z + \imath \beta)^{g - s} \\[6pt]
		& \times \int_0^1 du \; (1 - u)^{s - x - 1} \; u^{g+x - 1} \;\biggl( 1 - \biggl[ \frac{1}{2} + \frac{\imath z}{2\beta} \biggr] u \biggr)^{x - g}.
	\end{aligned}
\end{align}
This is exactly Euler representation for the hypergeometric function
\begin{align}
	\begin{aligned}
		\mathcal{K}(s,x) \cdot 1 & = \frac{\Gamma ( x + g)}{\Gamma ( s + g)}  \, \biggl( 1 - \biggl[ \frac{1}{2} + \frac{\imath z}{2\beta} \biggr] \biggr)^{g - s}  \\[6pt]
		& \times {}_2 F_1 \biggl( g - x , \; g+x, \; g+s; \; \frac{1}{2} + \frac{\imath z}{2\beta} \biggr),
	\end{aligned}
\end{align}
which converges provided that
\begin{align}
	\Re s > \Re x, \qquad \Re (x + g) > 0.
\end{align}
In particular, it converges in the case we consider in this paper
\begin{align}
	s > \frac{1}{2}, \qquad x = \imath \lambda \in \imath \mathbb{R}, \qquad
g > 0.
\end{align}
It is only left to use Euler transformation
\begin{align}
	(1 - w)^{a+b-c} {}_2 F_1 (a, b, c; w) = {}_2 F_1 (c-a, c- b, c; w)
\end{align}
to obtain desired formula \eqref{K2F1}.

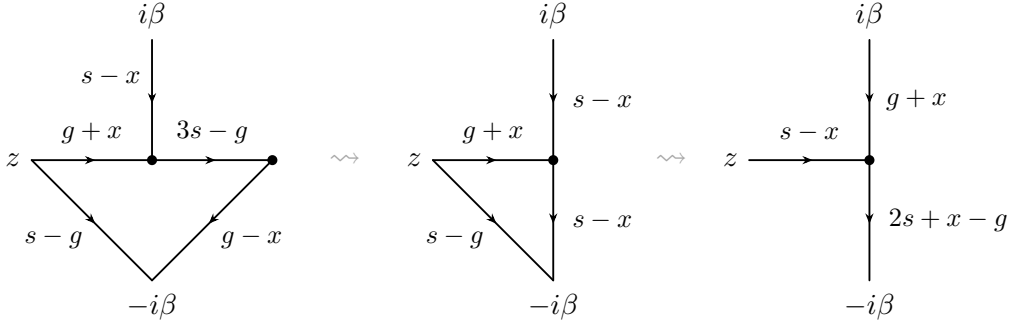
\begin{figure}[t]
	\centering
	\begin{tikzpicture}[thick, line cap = round]
		\def\l{1.6}
		\def\r{1.5pt}
		\draw[->-] (0,0) node[left] {\small $z$} to node[midway,above=2pt]{\footnotesize $g+x$} (\l,0);
		\draw[fill = black] (\l,0) circle (\r);
		\draw[->-] (\l,0) to node[midway,above=2pt]{\footnotesize $3s - g$} (2*\l,0);
		\draw[->-] (\l, \l) node[above]{\small $\imath \beta$} to node[pos = 0.3, left]{\footnotesize $s - x$} (\l, 0);
		\draw[->-] (0,0) to node[pos = 0.65, left = 6pt]{\footnotesize $s - g$} (\l, - \l);
		\draw[->-] (2*\l, 0) to node[pos = 0.65, right = 6pt]{\footnotesize $g-x$} (\l, -\l) node[below = 1pt]{\small $-\imath \beta$};
		\draw[fill = black] (2*\l,0) circle (\r) node[right]{\color{gray!70!white}{$\hspace{0.6cm}\rightsquigarrow \hspace{0.2cm}$}};
	\end{tikzpicture}
	\begin{tikzpicture}[thick, line cap = round]
		\def\l{1.6}
		\def\r{1.5pt}
		\draw[->-] (0,0) node[left] {\small $z$} to node[midway,above=2pt]{\footnotesize $g+x$} (\l,0);
		\draw[fill = black] (\l,0) circle (\r);
		\draw[->-] (\l,0) to node[midway,right=3pt]{\footnotesize $s-x$} (\l,-\l);
		\draw[->-] (\l, \l) node[above]{\small$\imath \beta$} to node[midway, right=3pt]{\footnotesize $s - x$} (\l, 0);
		\draw[->-] (0,0) to node[pos = 0.65, left = 6pt]{\footnotesize $s - g$} (\l, - \l) node[below = 1pt]{\small$-\imath \beta$};
	\end{tikzpicture}\hspace{-0.1cm}
	\begin{tikzpicture}[thick, line cap = round]
		\def\l{1.6}
		\def\r{1.5pt}
		\draw[->-] (0,0) node[left] {{\color{gray!70!white}$\rightsquigarrow$} \hspace{0.2cm} {\small$z$}} to node[midway,above=3pt]{\footnotesize $s - x $} (\l,0);
		\draw[fill = black] (\l,0) circle (\r);
		\draw[->-] (\l,0) to node[midway,right=3pt]{\footnotesize $2s + x - g$} (\l,-\l) node[below = 1pt]{\small$-\imath \beta$};
		\draw[->-] (\l, \l) node[above]{\small$\imath \beta$} to node[midway, right=2pt]{\footnotesize $g+x$} (\l, 0);
	\end{tikzpicture}
	\caption{Transforming $\mathcal{K}(s, x) \cdot 1$ into the hypergeometric function} \label{fig:K1}
\end{figure}

The same transformation of $\mathcal{K}(s, x) \cdot 1$ into the hypergeometric function in terms of diagrams is shown in Figure \ref{fig:K1}. First, we use chain relation (Figure~\ref{fig:chain}) and at the second step --- Euler transformation (Figure~\ref{fig:Euler}). The resulting diagram up to coefficient depicts the integral representation \eqref{F1}
\begin{multline}\label{F2}
		{}_2 F_1 \biggl(s+x ,  s-x ,  s + g ; \, \frac{1}{2} + \frac{\imath z}{2\beta} \biggr) =
		(2i\beta)^{s+x}\,e^{i\pi s}\,\frac{\Gamma(s + g)\Gamma(2s+x - g)}{\Gamma(s+x)\Gamma(2s)} \\[8pt]
		\times \int \mathcal{D} w \; \frac{1}{(z-\bar{w})^{s-x} \, (i\beta-\bar{w})^{g + x} \, (w+i\beta)^{2s+x - g}}
\end{multline}
derived in Appendix \ref{Diagrammar}.

\section{Orthogonality}\label{sec:orth}
Recall the scalar product \eqref{sc}
\begin{align}\label{sc2}
	\langle \chi| \psi\rangle = \int \mathcal{D} z \; \overline{\chi(z)} \, \psi(z),
\end{align}
where integration is performed over upper half-plane $\Im z > 0$ and the measure is given by
\begin{align}
	\mathcal{D} z = \frac{2s - 1}{\pi}\, (2 \Im z)^{2s - 2} \; d\Re z \; d \Im z.
\end{align}
In this section we prove in two ways that eigenfunctions
\begin{align}\label{psi2}
	\Psi_\lambda(z) =  {}_2 F_1 \biggl( s + \imath \lambda, \, s - \imath \lambda, \,  s + g ; \,  \frac{1}{2} + \frac{\imath z}{2\beta} \biggr)
\end{align}
with $\lambda \in \mathbb{R}$ are orthogonal with respect to this scalar product
\begin{align}\label{orth}
	\langle \Psi_\rho | \Psi_\lambda \rangle = \mu^{-1}(\lambda) \; \frac{\delta(\lambda - \rho) + \delta(\lambda + \rho)}{2}.
\end{align}
The normalization coefficient
\begin{align}\label{mu}
	\mu(\lambda) = \frac{1}{4\pi \, (2\beta)^{2s} \, \Gamma(2s)} \; \biggl| \frac{\Gamma^2(s + \imath \lambda) \, \Gamma ( g+\imath \lambda ) }{\Gamma( s + g ) \, \Gamma(2\imath \lambda)}\biggr|^2
\end{align}
plays the role of integration measure in completeness relation, see Section~\ref{sec:compl}. Note that eigenfunctions \eqref{psi2} are invariant under reflection~$\lambda \to -\lambda$, so that in the relation \eqref{orth} we have symmetrized delta function.

\subsection{Proof from asymptotics}\label{sec:orth1}
The first proof is standard and usually mentioned in introductory texts on quantum mechanics (for example, see \cite[\S 36]{FY}). The idea is to cut-off the scalar product
\begin{equation}
	\langle\chi| \psi\rangle = \lim_{R \to \infty} \langle \chi | \psi \rangle_R,
\end{equation}
so that we can write
\begin{align}\label{scR}
	\begin{aligned}
		\langle \Psi_\rho | \Psi_\lambda \rangle &= \lim_{R \to \infty} \langle \Psi_\rho | \Psi_\lambda \rangle_R \\[6pt]
		& = \lim_{R \to \infty} \frac{1}{\rho^2 - \lambda^2} \Bigl[ \langle \Psi_\rho |  H^s \Psi_\lambda\rangle_R - \langle H^s \Psi_\rho |  \Psi_\lambda\rangle_R \Bigr] \\[6pt]
		& = \lim_{R \to \infty} \frac{1}{\rho^2 - \lambda^2} \bigl[ \text{ boundary terms } \bigr],
	\end{aligned}
\end{align}
where on the last step we integrate by parts. Then we only need to calculate asymptotics of boundary terms. For this calculation we choose polar coordinates
\begin{equation}
	z = r e^{\imath \phi}
\end{equation}
and scalar product with radius cut-off
\begin{equation}
	\langle \chi |  \psi \rangle_R =  \frac{2s - 1}{\pi}  \int_{0}^\pi d\phi \, (2\sin \phi)^{2s - 2} \int_0^R dr \, r^{2s - 1} \, \overline{\chi(r e^{\imath \phi})} \; \psi(r e^{\imath \phi}).
\end{equation}
Moreover, since Hamiltonian \eqref{H} can be expressed in terms of spin operators
\begin{equation}\label{H3}
	H^{s} = S^2 + \beta^2 S_-^2 - 2\imath \alpha S_-, \qquad S = z\partial_z + s, \qquad S_- = - \partial_z,
\end{equation}
we first calculate boundary terms that arise from each spin operator (when we integrate it by parts).

Rewriting derivatives
\begin{align}
	\partial_z = \frac{e^{- \imath \phi}}{2} \partial_r + \frac{e^{-\imath \phi}}{2 \imath r} \partial_\phi, \qquad \partial_{\bar{z}} = \frac{e^{ \imath \phi}}{2} \partial_r - \frac{e^{\imath \phi}}{2 \imath r} \partial_\phi
\end{align}
we express action of spin operators on analytic functions as
\begin{align}\label{S2}
	& S \psi(z) = (z \partial_z + \bar{z} \partial_{\bar{z}} + s) \psi(z) = (r \partial_r + s) \psi(z), \\[6pt] \label{S-}
	& S_- \psi(z) = -(\partial_z + \partial_{\bar{z}}) \psi(z) = \biggl( \frac{\sin\phi}{r} \, \partial_\phi - \cos\phi \, \partial_r \biggr) \psi(z).
\end{align}
Next define Wronskians with respect to polar coordinates
\begin{align}
	& W_r[\chi, \psi] = \chi \, \partial_r \psi - \psi  \, \partial_r \chi, \\[6pt]
	& W_\phi[\chi, \psi]= \chi \, \partial_\phi \psi - \psi \, \partial_\phi \chi.
\end{align}
Using \eqref{S2}, \eqref{S-} and integrating by parts we establish the following two identities
\begin{align}
	& \langle \chi |  S \psi\rangle_R = \langle -S \chi | \psi\rangle_R + \frac{2s - 1}{\pi} \, R^{2s} \int_0^\pi d\phi \, (2 \sin\phi)^{2s - 2}  \;  \bigl( \bar{\chi} \, \psi \bigr) \bigr|_{r = R}, \\[10pt]
	& \label{Sm}
	\begin{aligned}
		\langle \chi |  S_- \psi\rangle_R &= \langle -S_- \chi | \psi \rangle_R \\[6pt]
		&- \frac{2s - 1}{\pi} \, R^{2s - 1} \int_0^\pi d\phi\, (2\sin \phi)^{2s - 2}  \, \cos \phi \;  \bigl( \bar{\chi} \, \psi \bigr) \bigr|_{r = R}.
	\end{aligned}
\end{align}
Notice that for the functions $\chi, \psi$ decaying fast enough boundary terms tend to zero as $R \to \infty$, so that spin operators are anti-hermitian with respect to original scalar product. Iterating the above identities two times we obtain
\begin{align}\label{S3}
	& \begin{aligned}
		\langle \chi | S^2 \psi\rangle_R &= \langle S^2 \chi | \psi \rangle_R \\[6pt]
		&+ \frac{2s - 1}{\pi} \, R^{2s + 1} \int_0^\pi d\phi \, (2\sin\phi)^{2s - 2}  \;  W_r[\bar{\chi}, \psi]  \bigr|_{r = R},
	\end{aligned} \\[10pt] \label{Sm2}
	& \begin{aligned}
		\langle \chi | S_-^2 \psi\rangle_R &= \langle S_-^2 \chi | \psi \rangle_R \\[6pt]
		&+ \frac{2s - 1}{\pi} \, R^{2s - 1} \int_0^\pi d\phi \, (2\sin \phi)^{2s - 2} \, \cos^2 \phi \; \, W_r[\bar{\chi}, \psi]\bigr|_{r = R}\\[6pt]
		&+ \frac{2s - 1}{2\pi} \, R^{2s - 2} \int_0^\pi d\phi \, (2\sin \phi)^{2s - 1} \, \cos \phi \; \, W_\phi[\psi, \bar{\chi}]\bigr|_{r = R}.
	\end{aligned}
\end{align}
Thus, when we integrate by parts the Hamiltonian \eqref{H3}
\begin{align}
	\langle \Psi_\rho | H^s \Psi_\lambda\rangle_R = \langle H^s \Psi_\rho | \Psi_\lambda\rangle_R \; + \text{ boundary terms},
\end{align}
boundary terms are given by the formulas \eqref{Sm}, \eqref{S3}, \eqref{Sm2}.

To calculate their asymptotics as $R \to \infty$ we use well-known formula for the hypergeometric function \cite[\href{http://dlmf.nist.gov/15.12.i}{Sec. 15.12(i)}]{DLMF}
\begin{equation}\label{F-as}
	{}_2F_1(a,b,c;w) \sim \frac{\Gamma(b - a)\Gamma(c)}{\Gamma(b) \Gamma(c - a)} (-w)^{-a} + \frac{\Gamma(a - b)\Gamma(c)}{\Gamma(a) \Gamma(c - b)} (-w)^{-b}
\end{equation}
as $|w| \to \infty$. From that for the eigenfunctions
\begin{equation}
	\Psi_\lambda(z) = {}_2 F_1 \biggl( s + \imath \lambda, \, s - \imath \lambda, \,  s + g ; \,  \frac{1}{2} + \frac{\imath z}{2\beta} \biggr)
\end{equation}
the needed asymptotics as $|z| \to \infty$ have the form
\begin{align}\label{Psi-as}
	&\Psi_{\lambda}(z) \sim c(\lambda) \, z^{-s + \imath \lambda} + c(-\lambda) \, z^{-s - \imath \lambda}, \\[6pt] \label{Psir-as}
	&\partial_r \Psi_{\lambda}(z) \sim  \frac{-s + \imath \lambda}{|z|} \, c(\lambda) \, z^{-s + \imath \lambda} + \frac{-s - \imath \lambda}{|z|} \, c(-\lambda) \, z^{-s - \imath \lambda} , \\[6pt]
	& \partial_\phi \Psi_{\lambda}(z) \sim \imath (-s + \imath \lambda) \, c(\lambda) \, z^{-s + \imath \lambda}  + \imath (-s - \imath \lambda) \, c(-\lambda) \, z^{-s - \imath \lambda},
\end{align}
where the coefficient
\begin{equation}\label{c}
	c(\lambda) = \frac{\Gamma(2 \imath \lambda) \, \Gamma ( s + g)}{\Gamma(s + \imath \lambda) \,
\Gamma ( g + \imath \lambda )} \; (2 \beta)^{s - \imath \lambda} \; e^{\frac{\imath \pi s}{2} + \frac{\pi \lambda}{2}}.
\end{equation}

\begin{remark} \label{rem2}
	In this paper we only consider the case $g > 0$. Note that if $g < 0$ the
coefficient $c(\lambda)$ \eqref{c} has zeroes
	\begin{equation}
		\lambda = \imath ( n + g), \qquad n \in \mathbb{N}_0
	\end{equation}
	in the lower half-plane $\Im \lambda < 0$, which may correspond to discrete spectra (as it can be seen from asymptotics~\eqref{Psi-as}).
\end{remark}

Substituting the above asymptotics into the boundary terms \eqref{Sm}, \eqref{S3}, \eqref{Sm2}, we conclude that the only boundary term that gives nonzero contribution as $R \to \infty$ is from~$S^2$~\eqref{S3}
\begin{align}\label{S4}
	\langle \Psi_\rho | \Psi_\lambda \rangle_R \sim \frac{R^{2s + 1}}{\rho^2 - \lambda^2}  \, \frac{2s - 1}{\pi} \int_0^\pi d\phi \, (2\sin\phi)^{2s - 2}  \;  W_r\bigl[ \overline{\Psi}_\rho, \Psi_\lambda\bigr]  \bigr|_{r = R}.
\end{align}
One could guess that without calculations just from the behaviour of spin operators \eqref{S2}, \eqref{S-} as $r \to \infty$.

From the above formulas \eqref{Psi-as}, \eqref{Psir-as} we deduce asymptotics of Wronskian
\begin{equation}
	W_r\bigl[ \overline{\Psi}_\rho, \Psi_\lambda\bigr] \sim W_{\lambda, \rho} + W_{-\lambda, -\rho} + W_{-\lambda, \rho} + W_{\lambda, -\rho} , \qquad |z| \to \infty,
\end{equation}
where we denoted
\begin{equation}
	W_{\lambda, \rho} = \frac{\imath (\lambda + \rho)}{|z|^{2s + 1}} \; \overline{c(\rho)} \, c(\lambda) \, \bar{z}^{- \imath \rho} \, z^{ \imath \lambda} .
\end{equation}
Inserting this into \eqref{S4} we obtain
\begin{equation}
	\begin{aligned}
		&\frac{R^{2s + 1}}{\rho^2 - \lambda^2}  \, \frac{2s - 1}{\pi}\int_0^\pi d\phi \, (2\sin\phi)^{2s - 2}  \;  W_r\bigl[ \overline{\Psi}_\rho, \Psi_\lambda\bigr]    \bigr|_{r = R} \\[12pt]
		&\hspace{1cm} \sim \, \overline{c(\rho)} \, c(\lambda) \, \frac{\imath R^{\imath (\lambda - \rho)}}{\rho - \lambda} \, \frac{2s - 1}{\pi} \int_0^\pi d\phi \, (2\sin \phi)^{2s - 2} \,  e^{-(\lambda + \rho) \phi} \\[12pt]
		&\hspace{1cm} +  \, \overline{c(-\rho)} \, c(-\lambda) \, \frac{\imath R^{\imath (\rho - \lambda)}}{\lambda - \rho} \, \frac{2s - 1}{\pi} \int_0^\pi d\phi \, (2\sin \phi)^{2s - 2} \,  e^{(\lambda + \rho) \phi} \\[12pt]
		&\hspace{1cm} + \, [ \dots \lambda \to - \lambda \dots ].
	\end{aligned}
\end{equation}
The remaining integrals are calculated using formula \cite[\href{http://dlmf.nist.gov/5.12.E6}{(5.12.6)}]{DLMF}
\begin{equation}\label{I}
	\frac{2s - 1}{\pi} \int_0^\pi d\phi \, (2\sin \phi)^{2s - 2} \, e^{2a \phi} =   e^{\pi a} \, \frac{ \Gamma(2s)}{\Gamma(s + \imath a) \, \Gamma(s - \imath a)}.
\end{equation}
We note in passing that it is related to the well-known Cauchy integral \cite[p. 48]{AAR} by the change of coordinate $\phi \to t = \ctg \phi$.

So, we can rewrite $R \to\infty$ asymptotics of the scalar product with cut-off~\eqref{S4} in the following form
\begin{equation}\label{W}
	\begin{aligned}
		& \langle \Psi_\rho | \Psi_\lambda \rangle_R \sim  \frac{\imath \, \Gamma(2s) }{\Gamma \Bigl(s + \frac{\imath (\lambda + \rho)}{2} \Bigr) \, \Gamma \Bigl(s - \frac{\imath (\lambda + \rho)}{2} \Bigr)} \\[15pt]
		& \quad \times \frac{1}{\rho - \lambda} \, \Bigl( e^{- \frac{\pi(\lambda + \rho)}{2}} \, \overline{c(\rho)} \, c(\lambda) \,  R^{\imath (\lambda - \rho)} - e^{\frac{\pi(\lambda + \rho)}{2}} \, \overline{c(-\rho)} \, c(-\lambda) \, R^{\imath (\rho - \lambda)} \Bigr) \\[15pt]
		& \quad + [ \dots \lambda \to - \lambda \dots ].
	\end{aligned}
\end{equation}
Notice that singularities at $\rho = \pm \lambda$ cancel since coefficient \eqref{c} satisfies
\begin{equation}\label{c-eq}
	e^{-\pi \lambda} \, | c(\lambda) |^2 = e^{\pi \lambda} \, |c(- \lambda)|^2.
\end{equation}
Hence, if we divide fast-oscillating exponents into two parts
\begin{align}
	R^{ \pm \imath(\lambda - \rho)} = \cos\bigl[ (\lambda - \rho) \ln R \bigr] \pm \imath \sin\bigl[ (\lambda - \rho) \ln R \bigr],
\end{align}
sum of the terms with cosines tends to zero as $R \to \infty$ due to Riemann--Lebesgue lemma. For the remaining terms we use well-known delta-sequence formula 
\begin{equation}
	\lim_{R \to \infty} \frac{\sin \bigl[ (\lambda - \rho) \ln R \bigr]}{\lambda - \rho} = \pi \delta(\lambda - \rho)
\end{equation}
together with the relation \eqref{c-eq} and expression for $c(\lambda)$ \eqref{c}. At the end we arrive at the result
\begin{equation}\label{sc-res}
	\begin{aligned}
		\langle \Psi_\rho | \Psi_\lambda \rangle &= 2 \pi \, (2\beta)^{2s} \; \Gamma(2s) \\[6pt]
		&\times \biggl|  \frac{\Gamma( s + g ) \, \Gamma(2\imath \lambda) }{\Gamma^2(s + \imath \lambda) \,
\Gamma( g+\imath \lambda )} \biggr|^2 \, \Bigl( \delta(\lambda - \rho) + \delta(\lambda + \rho) \Bigr).
	\end{aligned}
\end{equation}

\subsection{Diagrammatic calculation}\label{sec:orth2}

The second proof of orthogonality is straightforward: we use the integral representation \eqref{F2}
\begin{align}
	\begin{aligned}
		\Psi_\lambda(z) &=
		(2i\beta)^{s+\imath \lambda}\,e^{i\pi s}\,\frac{\Gamma(s + g)\Gamma(2s+ \imath \lambda - g)}{\Gamma(s+\imath \lambda)\Gamma(2s)} \\[8pt]
		&\times \int \mathcal{D} w \; \frac{1}{(z-\bar{w})^{s-\imath \lambda} \, (i\beta-\bar{w})^{g + \imath \lambda} \, (w+i\beta)^{2s+\imath \lambda - g}}
	\end{aligned}
\end{align}
and explicitly calculate the regularized scalar product between the eigenfunctions. The calculation relies on the diagram technique explained in Appendix~\ref{Diagrammar}.

\begin{figure}[t]
	\centering
	\begin{tikzpicture}[thick, line cap = round]
		\def\l{1.5}
		\def\r{1.5pt}
		\def\k{1.4}
		\draw[->-] (0,0) to node[midway,above=3pt]{\footnotesize $s - \imath \lambda$} (\k*\l,0);
		\draw[->-] (\k*\l, \l) node[above]{$\imath \beta$} to node[midway, right=3pt]{\footnotesize $\imath \lambda +g$} (\k*\l, 0);
		\draw[->-] (\k*\l,0) to node[midway,right=3pt]{\footnotesize $2s + \imath \lambda -g$} (\k*\l,-\l) node[below = 1pt]{$-\imath \beta$};
		\draw[->-] (-\k*\l,0) to node[midway,above=3pt]{\footnotesize $s + \imath \rho + \varepsilon$} (0,0);
		\draw[->-] (-\k*\l, \l) node[above]{$\imath \beta$} to node[midway, left=3pt]{\footnotesize $2s - \imath \rho -g + \varepsilon$} (-\k*\l, 0);
		\draw[->-] (-\k*\l,0) to node[midway,left=3pt]{\footnotesize $-\imath \rho +g - \varepsilon$} (-\k*\l,-\l) node[below = 1pt]{$-\imath \beta$};
		\draw[fill = black] (0,0) circle (\r);
		\draw[fill = black] (\k*\l,0) circle (\r);
		\draw[fill = black] (-\k*\l,0) circle (\r);
	\end{tikzpicture}
	\caption{The integral $A_1$} \label{fig:A1}
\end{figure}
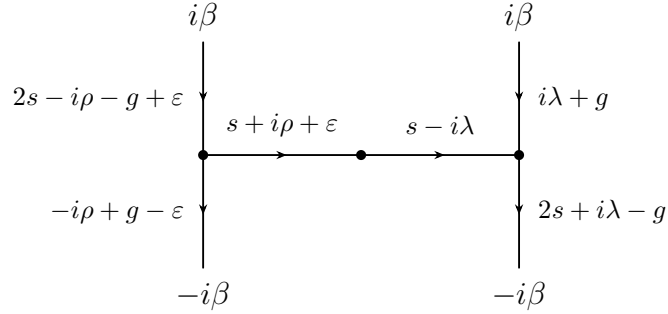
Substitution of the above representation into the scalar product results in expression
\begin{align}
	\begin{aligned}
		\langle \Psi_\rho | \Psi_\lambda \rangle &=
		\frac{\Gamma^2(s+g)
			\Gamma(2s+i\lambda-g)
			\Gamma(2s-i\rho-g)}
		{\Gamma^2(2s)\Gamma(s+i\lambda)\Gamma(s-i\rho)} \\[6pt]
		&\times  e^{2 \pi i s}  \, (2i\beta)^{s+i\lambda}\,
		(2i\beta)^{s-i\rho}\, \lim_{\varepsilon \to 0} A_1
	\end{aligned}
\end{align}
where the integral $A_1$ is depicted by the diagram in Figure \ref{fig:A1}. Since the scalar product is not absolutely convergent, we introduced regularization shifting some powers by $\varepsilon$.

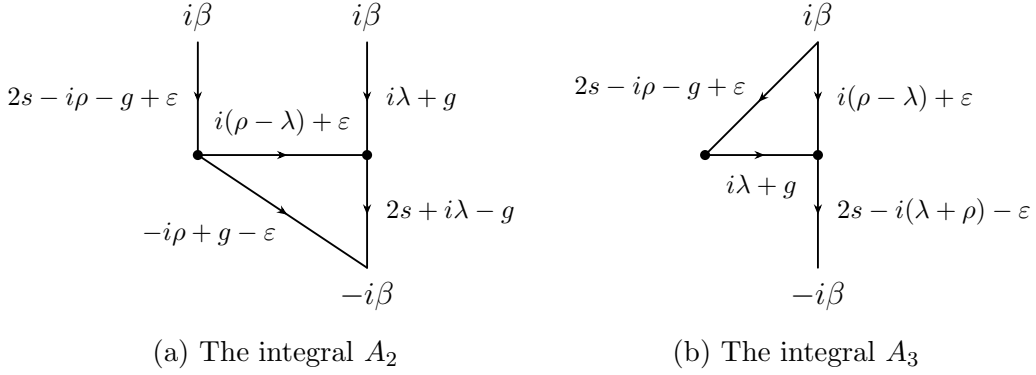
\begin{figure}[t]
	\centering
	\begin{subfigure}{0.54\textwidth}
	\begin{tikzpicture}[thick, line cap = round]
		\def\l{1.5}
		\def\k{1.5}
		\def\r{1.5pt}
		\draw[->-] (0,0) to node[midway,above=3pt]{\footnotesize $\imath (\rho - \lambda) + \varepsilon$} (\k*\l,0);
		\draw[->-] (\k*\l, \l) node[above]{$\imath \beta$} to node[midway, right=3pt]{\footnotesize $\imath \lambda +g$} (\k*\l, 0);
		\draw[->-] (\k*\l,0) to node[midway,right=3pt]{\footnotesize $2s + \imath \lambda -g$} (\k*\l,-\l) node[below = 1pt]{$-\imath \beta$};
		\draw[->-] (0, \l) node[above]{$\imath \beta$} to node[midway, left=3pt]{\footnotesize $2s - \imath \rho -g + \varepsilon$} (0, 0);
		\draw[->-] (0,0) to node[pos=0.7,left=11pt]{\footnotesize $-\imath \rho +g - \varepsilon$} (\k*\l,-\l);
		\draw[fill = black] (0,0) circle (\r);
		\draw[fill = black] (\k*\l,0) circle (\r);
	\end{tikzpicture}
		\caption{The integral $A_2$} \label{fig:A2}
	\end{subfigure}
	\begin{subfigure}{0.45\textwidth}
	\begin{tikzpicture}[thick, line cap = round]
		\def\l{1.5}
		\def\k{1}
		\def\r{1.5pt}
		\draw[->-] (0,0) to node[midway,below=3pt]{\footnotesize $\imath \lambda +g$} 	(\k*\l,0);
		\draw[->-] (\k*\l, \l) node[above]{$\imath \beta$} to node[midway, 	right=3pt]{\footnotesize $\imath (\rho - \lambda) + \epsilon$} (\k*\l, 0);
		\draw[->-] (\k*\l,0) to node[midway,right=3pt]{\footnotesize $2s - \imath( \lambda + 	\rho) - \epsilon$} (\k*\l,-\l) node[below = 1pt]{ $-\imath \beta$};
		\draw[->-] (\k*\l, \l) to node[pos=0.4, 	left=6pt]{\footnotesize $2s - \imath \rho -g + \epsilon$} (0, 0);
		\draw[fill = black] (0,0) circle (\r);
		\draw[fill = black] (\k*\l,0) circle (\r);
	\end{tikzpicture}
	\caption{The integral $A_3$} \label{fig:A3}
\end{subfigure}
\caption{Transformations of the scalar product}
\end{figure}
\begin{figure}[t]
	\centering
	\begin{tikzpicture}[thick, line cap = round]
		\def\l{1.5}
		\def\k{1}
		\def\r{1.5pt}
		\draw[->-] (0,0) to node[midway,below=3pt]{\scriptsize $\imath \lambda +g$} (\k*\l,0);
		\draw[->-] (\k*\l, \l) node[above]{\small$\imath \beta$} to node[midway, right=3pt]{\scriptsize $\imath (\rho - \lambda) + \epsilon$} (\k*\l, 0);
		\draw[->-] (\k*\l,0) to node[midway,right=3pt]{\scriptsize $2s - \imath( \lambda + \rho) - \epsilon$} (\k*\l,-\l) node[below = 1pt]{\small $-\imath \beta$};
		\draw[->-] (\k*\l, \l) to node[pos=0.4, left=6pt]{\scriptsize $2s - \imath \rho -g + \epsilon$} (0, 0);
		\draw[fill = black] (0,0) circle (\r);
		\draw[fill = black] (\k*\l,0) circle (\r);
	\end{tikzpicture}
	\begin{tikzpicture}[thick, line cap = round]
		\def\l{1.5}
		\def\k{1.2}
		\def\r{1.5pt}
		\def\d{3.7}
		\draw[->-] (\k*\l, \l) node[above]{\small$\imath \beta$} to node[midway, right=3pt]{\scriptsize $2\epsilon$} (\k*\l, 0);
		\draw[->-] (\k*\l,0) to node[midway,right=3pt]{\scriptsize $2s - \imath( \lambda + \rho) - \epsilon$} (\k*\l,-\l) node[below = 1pt]{\small$-\imath \beta$};
		\draw[fill = black] (\k*\l,0) circle (\r) node[left = 0.5cm]{\color{gray!70!white}$ \rightsquigarrow \; \;$};
		
		\node (z) at (\d+1,0) {\color{gray!70!white}$ \rightsquigarrow \; \;$};
		\draw[->-] (\d+\k*\l,0.5*\l) node[above]{\small$\imath \beta$} to node[midway,right=3pt]{\scriptsize $\epsilon - \imath( \lambda + 	\rho)$} (\d+\k*\l,-0.5*\l) node[below = 1pt]{\small$-\imath \beta$};
	\end{tikzpicture}
	\caption{Calculation of the integral $A_3$} \label{fig:A3-2}
\end{figure}
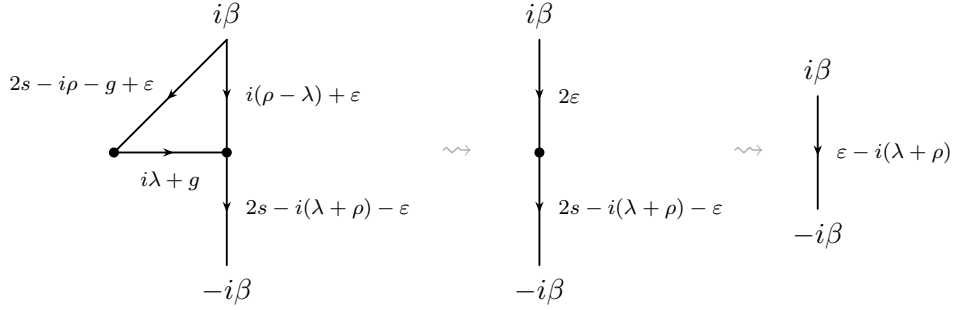

Next we use chain rule (see Figure \ref{fig:chain}) for two lines in the middle and obtain the diagram in Figure \ref{fig:A2}
\begin{align}
A_1 = e^{-i\pi s} \frac{\Gamma(i(\rho-\lambda)+\varepsilon)\Gamma(2s)}
{\Gamma(s+i\rho+\varepsilon)\Gamma(s-i\lambda)}\, A_2.
\end{align}
After that we perform Euler transformation (Figure \ref{fig:Euler}) of the right vertex in Figure \ref{fig:A2}. The resulting integral
\begin{align}
A_2 = (2i\beta)^{-2i\lambda}\,\frac{\Gamma(2s-i(\rho+\lambda)-\varepsilon)\Gamma(i(\rho+\lambda)+\varepsilon)}
{\Gamma(2s-g+i\lambda)
\Gamma(g-i\lambda)}\, A_3
\end{align}
is depicted in Figure \ref{fig:A3}. It remains to perform two integrations using chain rules, as shown in Figure \ref{fig:A3-2}, which gives
\begin{align}
	\begin{aligned}
		A_3 &= e^{-i\pi s} \frac{\Gamma(i(\lambda-\rho)+\varepsilon)\Gamma(2s)}
		{\Gamma\left(2s-g-i\rho +\varepsilon\right)
			\Gamma\left(g+i\lambda\right)} \\[6pt]
		&\times e^{-i\pi s} \frac{\Gamma(-i(\lambda+\rho)+\varepsilon)\Gamma(2s)}
		{\Gamma(2s-i(\rho+\lambda) - \varepsilon)
			\Gamma(2\varepsilon)}\,(2i\beta)^{i(\rho+\lambda) - \varepsilon}.
	\end{aligned}
\end{align}
\noindent Collecting everything together we obtain
\begin{align}
	\begin{aligned}
		\langle \Psi_\rho | \Psi_\lambda \rangle &= (2\beta)^{2s} \,
		\frac{\Gamma^2(g+s)\Gamma(2s)}
		{\Gamma(g\pm i\lambda)
			\Gamma(s\pm i\lambda)\Gamma(s\pm i\rho)}\,
		 \\[6pt]
		&\times\lim_{\varepsilon \to 0} \frac{\Gamma(\pm i(\lambda+\rho)+\varepsilon)
			\Gamma(\pm i(\lambda-\rho)+\varepsilon)}{\Gamma(2\varepsilon)},
	\end{aligned}
\end{align}
where for brevity we denoted
\begin{equation}
	f(a \pm b) = f(a + b) f(a - b).
\end{equation}
The remaining limit can be calculated using
Sokhotski-Plemelj theorem
\begin{align}
\lim_{\varepsilon \to 0} \frac{2\varepsilon}{(x\pm i\varepsilon)} = \lim_{\epsilon \to 0} \biggl( \frac{\imath }{x + \imath \epsilon} - \frac{\imath}{x - \imath \epsilon} \biggr) = 2\pi \delta(x)
\end{align}
that gives
\begin{multline}
\lim_{\varepsilon \to 0} \frac{\Gamma(\pm i(\lambda+\rho)+\varepsilon)
\Gamma(\pm i(\lambda-\rho)+\varepsilon)}{\Gamma(2\varepsilon)} \\[4pt]
= 2\pi \Gamma(\pm 2i\lambda)\,\Bigl( \delta(\lambda - \rho) + \delta(\lambda + \rho) \Bigr).
\end{multline}
Finally, we obtain the same result as in the previous section \eqref{sc-res}
\begin{multline}
\langle \Psi_\rho | \Psi_\lambda \rangle =
2\pi\, (2\beta)^{2s}\, \frac{\Gamma(2s) \, \Gamma^2(g+s) \,
\Gamma(\pm 2i\lambda)}
{\Gamma(g\pm i\lambda) \,
\Gamma^2(s\pm i\lambda)}\,
\Bigl( \delta(\lambda - \rho) + \delta(\lambda + \rho) \Bigr).
\end{multline}

\section{Completeness}\label{sec:compl}

In this section we prove the completeness relation
\begin{equation}\label{comp}
	I(z, \bar{w}) = \int_\mathbb{R} d\lambda \; \mu(\lambda)  \; \Psi_\lambda(z) \, \overline{\Psi_\lambda(w)}= \frac{e^{\imath \pi s}}{(z - \bar{w})^{2s}},
\end{equation}
where the integration measure
\begin{align}\label{mu-2}
	\mu(\lambda) = \frac{1}{4\pi  \, (2 \beta)^{2s} \, \Gamma(2s)} \; \biggl| \frac{\Gamma^2(s + \imath \lambda) \,
\Gamma ( g+\imath \lambda ) }{ \Gamma ( s + g) \,  \Gamma(2\imath \lambda)}\biggr|^2
\end{align}
follows from the orthogonality relation \eqref{orth}. From the right in \eqref{comp} we have the kernel of identity operator, that is
\begin{equation}\label{repr}
	\psi(z) = \int \mathcal{D}w \; \frac{e^{\imath \pi s}}{(z - \bar{w})^{2s}} \; \psi(w)
\end{equation}
for any $\psi(z)$ analytic in the upper half-plane \cite[(A.7)]{DKM1}. The strategy of the proof is as follows: first, we calculate the integral \eqref{comp} assuming
\begin{align}\label{zw}
	\Im z > \beta, \qquad \Im w > \beta.
\end{align}
Secondly, we show that $I(z, \bar{w})$ is absolutely convergent uniformly in $z, w$ from the upper half-plane, so that we can analytically continue the answer.

To calculate the integral \eqref{comp} we use Barnes integral representation of the hypergeometric function
\begin{align}\label{B}
	\Psi_\lambda(z) = \frac{\Gamma ( s+g )}
{\Gamma(s \pm \imath \lambda)} \int_{\mathbb{R} + \imath c} \frac{d\rho}{2\pi} \; \frac{\Gamma(s \pm \imath \lambda + \imath \rho) \, \Gamma(- \imath \rho)}{\Gamma ( s + g + \imath \rho )} \; \biggl( - \frac{\imath z}{2\beta}  - \frac{1}{2} \biggr)^{\imath \rho}.
\end{align}
Here and in what follows denote for brevity
\begin{equation}
	f(a \pm b) = f(a + b) f(a - b).
\end{equation}
The horizontal contour should separate series of poles
\begin{align}
	\rho = -i n, \qquad \rho = \pm \lambda + i (n + s), \qquad n \in \mathbb{N}_0,
\end{align}
hence, we choose $c \in (0, s)$.

For the proof we also need two integrals with gamma functions. The first one is simple \cite[\href{http://dlmf.nist.gov/5.13.E1}{(5.13.1)}]{DLMF}
\begin{equation}\label{I1}
	\int_\mathbb{R} d\lambda \; \Gamma(a + \imath \lambda) \, \Gamma(b - \imath \lambda) \, z^{\imath \lambda} = 2\pi \, \Gamma(a + b) \, \frac{z^b}{(1 + z)^{a+b}} .
\end{equation}
Here we assume $\Re a$, $\Re b > 0$, so that the contour $\mathbb{R}$ separates series of poles
\begin{equation}
	\lambda = \imath (a + n), \qquad \lambda = - \imath (b + n), \qquad n \in \mathbb{N}_0.
\end{equation}
This integral can be calculated by evaluating residues or using Mellin transform \cite[Section 3.3.4]{PK}.

The second needed integral
\begin{multline}\label{I2}
	\int_\mathbb{R} d\lambda \; \frac{\Gamma(a_1 \pm \imath \lambda) \, \Gamma(a_2 \pm \imath \lambda) \, \Gamma(a_3  \pm \imath \lambda)}{\Gamma(\pm 2\imath \lambda)} \\[6pt]
	= 4\pi \, \Gamma(a_1 + a_2) \, \Gamma(a_1 + a_3) \, \Gamma(a_2 + a_3)
\end{multline}
is the limiting case of de Branges--Wilson integral \cite[\href{http://dlmf.nist.gov/5.13.E5}{(5.13.5)}]{DLMF} when one of the parameters $a_4$ is sent to $\infty$. Here we assume $\Re a_j > 0$ and the contour~$\mathbb{R}$ separates series of poles
\begin{equation}
	\lambda  = \pm \imath (a_j + n), \qquad j = 1,2,3, \qquad n\in \mathbb{N}_0.
\end{equation}
We note in passing that de Branges--Wilson integral coincides with $\mathrm{sp}(1)$ Gustafson integral \cite[Theorem 9.3]{G}.

Now insert Barnes representation into the integral \eqref{I}
\begin{equation}\label{I-3int}
	\begin{aligned}
		I = C \, \int_{\mathbb{R}}  d\lambda \; \frac{ \Gamma ( g\pm \imath \lambda )}{ \Gamma(\pm 2 \imath \lambda)} \; & \int\limits_{\mathbb{R} + \imath c} d\rho \; \frac{\Gamma(s \pm \imath \lambda + \imath \rho) \, \Gamma(- \imath \rho)}{\Gamma ( s + g + \imath \rho )} \; \biggl( - \frac{\imath z}{2\beta} - \frac{1}{2} \biggr)^{\imath \rho} \\[6pt]
		& \hspace{-1cm} \times \int\limits_{\mathbb{R} - \imath c} d\nu \; \frac{\Gamma(s \pm \imath \lambda - \imath \nu) \, \Gamma(\imath \nu)}{\Gamma ( s + g - \imath \nu )} \; \biggl( \frac{\imath \bar{w}}{2\beta} - \frac{1}{2} \biggr)^{-\imath \nu}.
	\end{aligned}
\end{equation}
All constants are hidden in the prefactor
\begin{equation}
	C =  \frac{1}{16  \pi^3 \, (2\beta)^{2s} \, \Gamma(2s)}.
\end{equation}
The above multiple integral \eqref{I-3int} is absolutely convergent under assumptions~\eqref{zw}. To prove it use the reflection formula
\begin{align}
	\Gamma( 2 i \lambda) \, \Gamma(- 2 i \lambda) = \frac{\pi}{2 \lambda \, \sh (2\pi\lambda )}
\end{align}
and the bounds on gamma functions 
\begin{align}
	& | \Gamma(a + i b) | \leq \sqrt{2\pi} \, e^{\frac{1}{6a}} \, |a + i b|^{a - \frac{1}{2}} \, e^{- \frac{\pi}{2} |b|}, \\[6pt]
	& \bigl| \Gamma^{-1}(a + i b) \bigr| \leq \frac{e^{a + \frac{1}{6a}} }{\sqrt{2\pi}} \, |a + i b|^{-a +\frac{1}{2}} \, e^{\frac{\pi}{2} |b|},
\end{align}
where $a > 0$, $b \in \mathbb{R}$. These bounds follow from \cite[p. 34]{PK}. Therefore, we can change the order of integrations and first integrate over~$\lambda$. 

The integral over $\lambda$ is the reduction of de Branges--Wilson integral \eqref{I2}. The contour $\mathbb{R}$ separates series of poles due to our assumption $g > 0$. This assumption corresponds to the case with no discrete spectra (see Remark~\ref{rem2}). Inserting the answer for this integral we obtain
\begin{equation}
	\begin{aligned}
		I = 4\pi C\, & \int\limits_{\mathbb{R} - \imath c} d\nu \; \Gamma(\imath \nu) \; \biggl( \frac{\imath \bar{w}}{2\beta} - \frac{1}{2} \biggr)^{-\imath \nu} \\[6pt]
		&\times \int\limits_{\mathbb{R} + \imath c} d\rho \; \Gamma(2s - \imath \nu + \imath \rho) \, \Gamma(- \imath \rho) \; \biggl( - \frac{\imath z}{2\beta} - \frac{1}{2} \biggr)^{\imath \rho}  .
	\end{aligned}
\end{equation}
The integral over $\rho$ is of the first type \eqref{I1} (up to shift $\rho = \rho' + \imath c$). Therefore \vspace{0.1cm}
\begin{equation}
		I = 8\pi^2 C  \int\limits_{\mathbb{R} - \imath c} d\nu \; \Gamma(\imath \nu) \;  \Gamma(2s - \imath \nu ) \;  \biggl( \frac{\imath \bar{w}}{2\beta} - \frac{1}{2} \biggr)^{-\imath \nu} \biggl( - \frac{\imath z}{2\beta} + \frac{1}{2} \biggr)^{\imath \nu - 2s}  .
\end{equation}
Here once again we have integral of the first type \eqref{I1}. Thus, substituting the answer for it we arrive at expected formula \eqref{comp}.

Now let us show that the integral $I(z, \bar{w})$ in initial form \eqref{comp} is absolutely convergent uniformly in $z, w$ varying on compact subsets of upper half-plane
\begin{align}
	\Im z > 0, \qquad \Im w > 0.
\end{align}
Since  the integrand is symmetric with respect to $\lambda \to -\lambda$, it is enough to check its asymptotics as $\lambda \to \infty$. For the measure \eqref{mu-2} using well-known asymptotics of gamma functions \cite[\href{http://dlmf.nist.gov/5.11.E9}{(5.11.9)}]{DLMF}
\begin{align}
	| \Gamma(a + i \lambda) | \sim \sqrt{2\pi} \, \lambda^{a - \frac{1}{2}} \, e^{- \frac{\pi}{2} \lambda}, \qquad \lambda \to \infty
\end{align}
we obtain
\begin{align}
	|\mu(\lambda)| \sim M \, \lambda^{4s + 2g - 2} \, e^{- \pi \lambda}
\end{align}
with some constant $M$. For the eigenfunctions in the case
\begin{align}\label{z-res}
	\Im z >0, \qquad z \not\in i(0, \beta)
\end{align}
from \cite[Theorem 3.1]{KD} we obtain
\begin{align}
	| \Psi_\lambda(z) | \sim f(z) \, \lambda^{\frac{1}{2} - s - g} \, e^{\lambda | \Im \zeta|}, \qquad \lambda \to  \infty,
\end{align}
where
\begin{equation}
	\zeta = \ln \Biggl( \frac{z}{i \beta} + \sqrt{-\frac{z^2}{\beta^2} - 1} \, \Biggr), \qquad z = i \beta \ch \zeta.
\end{equation}
Importantly, since $\Im z > 0$ we have
\begin{equation}
	| \Im \zeta | < \frac{\pi}{2}.
\end{equation}
In the complementary to \eqref{z-res} case
\begin{align}\label{z-restr}
	\Im z > 0, \qquad z \not\in i(\beta, \infty)
\end{align}
from \cite[Theorem 3.2]{KD} we get
\begin{align}
	| \Psi_\lambda(z) | \sim \tilde{f}(z) \, \lambda^{\frac{1}{2} - s - g} \, e^{\lambda (\pi + \Im \xi)}, \qquad \lambda \to  \infty,
\end{align}
where
\begin{align}
	\xi = \ln \Biggl( - \frac{z}{i \beta} - i \sqrt{ 1 + \frac{z^2}{\beta^2} } \,  \Biggr), \qquad z = - i \beta \ch \xi.
\end{align}
Here notice that due to restrictions \eqref{z-restr} we have
\begin{align}
	- \pi \leq \Im \xi < - \frac{\pi}{2}.
\end{align}
From the above asymptotics we conclude that for any $z, w$ from compact subsets of the upper half-plane
\begin{align}
	| \mu(\lambda) \, \Psi_\lambda(w) \, \Psi_\lambda(z) | \sim A(z, w) \, \lambda^{2s-1} \, e^{- \delta \lambda}, \qquad \lambda \to \infty,
\end{align}
with some $\delta > 0$. Thus, $I(z, \bar{w})$ is analytic function of $z, \bar{w}$ and we can remove the assumption \eqref{zw}.

\section{Relation to index hypergeometric transform}\label{sec:ind}

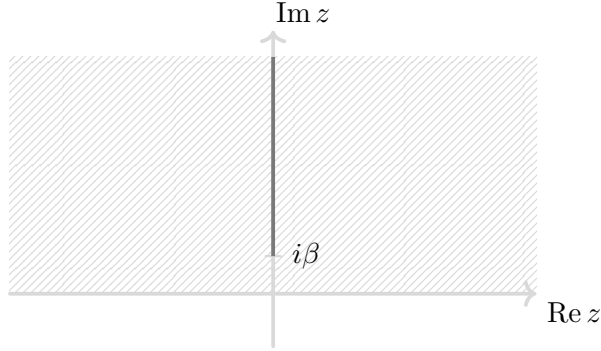
\begin{figure}[t]
	\centering
	\begin{tikzpicture}[line width=0.5mm]
		
		\def\l{3.5}
		\def\h{3.5}
		\def\b{0.5}
		\def\t{0.1}
		\def\n{0.42}
		
		\fill[pattern=north east lines, pattern color = gray!30!white] (-\l, 0) rectangle (\l, 0.9*\h);
		
		\draw[line cap = round, gray!30!white, ->] (-\l,0) -- (\l, 0) node[xshift = 0.5cm, yshift = -0.25cm] {\small \textcolor{black}{$\Re z$}};	
		\draw[line cap = round,gray!30!white, ->] (0, -0.2*\h) -- (0, \h) node[xshift = 0.4cm, yshift = 0.25cm] {\small \textcolor{black}{$\Im z$}};
		
		\draw[line cap = round, line width = 0.3mm, gray!30!white] (-\t, \b) -- (\t, \b) node[right] {\small \textcolor{black}{$\imath \beta$}};
		\draw[line width = 0.5mm, gray!95!black] (0, \b) -- (0, 0.9*\h);
	\end{tikzpicture}
	\caption{Upper half-plane $\Im z > 0$ and half-line $y > 0$}
	\label{fig:zx}
\end{figure}

The Hamiltonian \eqref{H} in terms of rescaled and shifted coordinate
\begin{align}
	y = -\frac{1}{2}  - \frac{\imath z}{2\beta}, \qquad z = \imath \beta (1 + 2y)
\end{align}
coincides with the standard hypergeometric operator
\begin{align}\nonumber
		H^s &= (z^2 + \beta^2) \partial_z^2 + (2s + 1) z \partial_z + 2\imath \alpha \partial_z + s^2 \\[10pt] \label{Hx}
		& = y (1 + y) \partial_y^2 + \bigl[ ( s + g ) + (2s + 1)y \bigr] \partial_y + s^2,
\end{align}
where as before $g = 1/2 + \alpha/\beta$. In addition to the scalar product over the upper half-plane $\Im z > 0$ \eqref{sc}
\begin{align}\label{scz}
	\langle \chi | \psi \rangle = \int \mathcal{D} z \; \overline{\chi(z)} \, \psi(z),
\end{align}
the operator \eqref{Hx} is also self-adjoint with respect to scalar product over the half-line $y > 0$, see Figure \ref{fig:zx},
\begin{align}\label{scx}
	( \chi | \psi) = \int_0^\infty dy \, \, y^{s + g - 1} \, (1 + y)^{s - g} \; \overline{\chi(y)} \, \psi(y).
\end{align}

\noindent In the previous sections we proved that eigenfunctions
\begin{align}
	\Psi_\lambda(z) = {}_2 F_1 \biggl( s + \imath \lambda, s - \imath \lambda, s + g; \, \frac{1}{2} + \frac{\imath z}{2\beta} \biggr)
\end{align}
with $\lambda \in \mathbb{R}$ are orthogonal and complete with respect to the first scalar product \eqref{scz}. Furthermore, they are also orthogonal and complete as functions of $y$
\begin{equation}
	\Phi_\lambda(y) \equiv \Psi_\lambda \bigl( \imath \beta(1 + 2y) \bigr)= {}_2 F_1 ( s + \imath \lambda, s - \imath \lambda, s + g ; \, -y )
\end{equation}
with respect to the second scalar product \eqref{scx}. In other words the corresponding transform $J$, which is known as index hypergeometric (or Jacobi) transform,
\begin{equation}\label{Jind}
	[ J \psi ] (\lambda) = \int_0^\infty dy \,\, y^{s + g - 1} \, (1 + y)^{s - g} \; \Phi_\lambda(y) \; \psi(y)
\end{equation}
is unitary, see the review article \cite{N}. Notice that $\overline{\Phi_\lambda(y)}$ = $\Phi_\lambda(y)$ for $y, \lambda \in \mathbb{R}$.

The eigenfunctions in two regimes $\Im z > 0$ and $y > 0$ are related by the following formulas
\begin{align} \label{Phi-Psi}
	& \Phi_\lambda(y) = \int \mathcal{D}z \; \frac{e^{\imath \pi s}}{\bigl(\imath \beta (1 + 2y) - \bar{z}\bigr)^{2s}} \; \Psi_\lambda(z), \\[15pt] \label{Psi-Phi}
	& \begin{aligned}
		\Psi_\lambda(z) = \frac{(2\beta)^{2s} \, \Gamma(2s)}{\Gamma(s + \imath \lambda) \, \Gamma(s - \imath \lambda)} \, \int_0^\infty & dy \,\, y^{s + g - 1} \, (1 + y)^{s - g} \\[6pt]
		& \times \frac{e^{\imath \pi s}}{\bigl( z + \imath \beta(1 + 2y) \bigr)^{2s}} \; \Phi_\lambda(y).
	\end{aligned}
\end{align}
In the first formula we just use the identity operator for the functions analytic in the upper half-plane \eqref{Rep}. The proof of the second formula is given in Appendix~\ref{App:Ind}.

Note that in the second formula \eqref{Psi-Phi} we essentially have the same kernel of identity operator. The kernel function $(z - \bar{w})^{-2s}$ appearing in both formulas satisfies 
\begin{align}\label{Hker}
	H^s(z) \, (z - \bar{w})^{-2s} = \overline{H^s(w)} \, (z - \bar{w})^{-2s}.
\end{align}
To prove it recall that the Hamiltonian can be written in terms of spin operators
\begin{align}
	H^s = S^2 + \beta^2 S_-^2 - 2 \imath \alpha S_-, \qquad S = z \partial_z + s, \qquad S_- = - \partial_z,
\end{align}
so that the formula \eqref{Hker} follows from the same type of relations for spin operators
\begin{equation}
	\begin{aligned}
		& S_a(z) \, (z - \bar{w})^{-2s} = -\overline{S_a(w)} \, (z - \bar{w})^{-2s}.
	\end{aligned}
\end{equation}
Therefore, the existence of the transformations \eqref{Phi-Psi} and \eqref{Psi-Phi} can be guessed from the identity \eqref{Hker} and self-adjointness of $H^s$ with respect to both scalar products \eqref{scz}, \eqref{scx}. However, these arguments doesn't fix the coefficients in these transformations.

Using formulas \eqref{Phi-Psi}, \eqref{Psi-Phi} we can also relate scalar products between the eigenfunctions. Denote transform between two Hilbert spaces corresponding to scalar products \eqref{scz}, \eqref{scx}
\begin{align}\label{U}
	[U  \psi] (y) = \int \mathcal{D}z \; \frac{e^{\imath \pi s}}{ \bigl(\imath \beta(1 + 2y) - \bar{z} \bigr)^{2s} } \; \psi(z).
\end{align}
Then the adjoint operator
\begin{align}\label{Uadj}
	[U^\dagger \chi] (z) = \int_0^\infty dy \,\, y^{s +g-1} \, (1 + y)^{s - g} \;
\frac{e^{\imath \pi s}}{\bigl( z + \imath \beta(1 + 2y) \bigr)^{2s}} \; \chi(y).
\end{align}
In these notations the formulas \eqref{Phi-Psi}, \eqref{Psi-Phi} read
\begin{align}
	\Phi_\lambda = U \, \Psi_\lambda, \qquad \Psi_\lambda = \frac{(2\beta)^{2s} \, \Gamma(2s)}{\Gamma(s + \imath \lambda) \, \Gamma(s - \imath \lambda)} \, U^\dagger \Phi_\lambda.
\end{align}
As a corollary, we have
\begin{align}
	U^\dagger U \, \Psi_{\lambda} = \frac{\Gamma(s + \imath \lambda) \, \Gamma(s - \imath \lambda)}{(2\beta)^{2s} \, \Gamma(2s)} \, \Psi_\lambda.
\end{align}
Using the last relation we connect scalar products between two families of orthogonal eigenfunctions
\begin{align}\label{scxz}
		( \Phi_\rho | \Phi_\lambda ) = ( U  \Psi_\rho | U  \Psi_\lambda ) = \frac{\Gamma(s + \imath \lambda) \, \Gamma(s - \imath \lambda)}{(2\beta)^{2s} \, \Gamma(2s)} \, \langle \Psi_\rho | \Psi_\lambda \rangle.
\end{align}
Similar relation between upper half-plane and half-line scalar products has been previously observed in the context of spin chains with many particles \cite[Section 6]{BDM}.

The formula \eqref{scxz} agrees with the known orthogonality and completeness relations for the index hypergeometric transform
\begin{align}
	& \int_0^\infty dy \,\, m(y) \, \Phi_\lambda(y) \, \overline{\Phi_\rho(y)} = \hat{\mu}^{-1}(\lambda) \, \frac{\delta(\lambda - \rho) + \delta(\lambda + \rho)}{2}, \\[6pt]
	& \int_\mathbb{R} d\lambda \,\, \hat{\mu}(\lambda) \, \Phi_\lambda(x) \, \overline{\Phi_\lambda(y)} = m^{-1}(y) \, \delta(x - y),
\end{align}
where for brevity we denoted measures 
\begin{align}
	m(y) = y^{s + g -1} \, (1 + y)^{s - g},\qquad \hat{\mu}(\lambda) = \frac{1}{4\pi} \biggl| \frac{\Gamma(s + i \lambda) \, \Gamma(g + i \lambda)}{\Gamma(s + g) \, \Gamma(2i \lambda) }\biggr|^2.
\end{align}
For example, compare with \cite[eq. (3.2), (4.5)]{K}.

The orthogonality and completeness of the eigenfunctions $\Psi_\lambda(z)$ is equivalent to the unitarity of the transform
\begin{align}
	[ T \psi ] (\lambda) = \int \mathcal{D}z \, \overline{\Psi_\lambda(z)} \, \psi(z).
\end{align}
Notice that it is related to the index hypergeometric transform $J$ \eqref{Jind} with the help of the same operator \eqref{Uadj}
\begin{equation}
	J = T U^\dagger.
\end{equation}

\subsection*{Acknowledgments}

The work of P. Antonenko (Sections~1,~2,~3) was supported by the Ministry of Science and Higher Education of the Russian Federation, agreement 075-15-2022-289, date 06/04/2022. The work of N. Belousov and S. Derkachov (Sections 4, 5, Appendices~A,~B) was supported by the Theoretical Physics and Mathematics Advancement Foundation BASIS. The work of S.~Khoroshkin (Section 6, Appendix C) was supported by Russian Science Foundation, project No. 23-11-00150.

\appendix

\renewcommand{\theequation}{\Alph{section}.\arabic{equation}}
\renewcommand{\thetable}{\Alph{table}}
\setcounter{section}{0}
\setcounter{table}{0}

\section{Diagram technique}\label{Diagrammar}

\begin{figure}[h]
	\centering
	\begin{tikzpicture}[thick, line cap = round]
		\def\l{2}
		\def\r{1.5pt}
		\draw[->-] (0,0) node[left]{$z$} to node[midway,above=3pt]{\footnotesize $\lambda$} (\l, 0) node[right]{$w \; = \; (z-\bar{w})^{-\lambda}$};
		\draw[fill = black] (7,0) circle (\r) node[xshift = 1.3cm] {$\displaystyle z \,\; = \; \int \mathcal{D}z$};
	\end{tikzpicture}
	\caption{Elements of diagrams} \label{fig:line}
\end{figure}

\noindent In this section we explain the diagram technique, which is used throughout the paper. The diagrams represent various integrals and consist of directed lines, which depict functions $(z-\bar{w})^{-\lambda}$, and bold vertices, which correspond to integration over upper half-plane $\Im z > 0$ with the measure
\begin{align}
	\mathcal{D}z = \frac{2s - 1}{\pi} \, (2\Im z)^{2s - 2} \; d \Re z \; d \Im z.
\end{align}

\begin{figure}[t]
	\centering
	\begin{tikzpicture}[thick, line cap = round]
		\def\l{2.2}
		\def\d{7}
		\def\r{1.5pt}
		\def\k{0.8}
		\draw[->-] (0,0) node[left] {$z$} to node[midway,above=3pt]{\footnotesize $\lambda$} (\k*\l,0);
		\draw[fill = black] (\k*\l,0) circle (\r);
		\draw[->-] (\k*\l,0) to node[midway,above=3pt]{\footnotesize $\rho$} (\k*2*\l,0) node[right]{$w \;\, = \,\; a(\lambda, \rho)$};
		\draw[->-] (\d,0) node[left]{$z$} to node[midway,above=3pt]{\footnotesize $\lambda + \rho - 2s$} (\d + \l,0) node[right]{$w$};
	\end{tikzpicture}
	\caption{Chain rule} \label{fig:chain}
\end{figure}
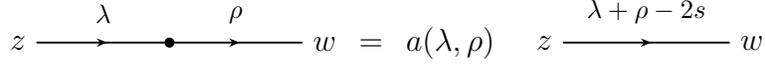

In the present work we use two diagram identities. The first one is the \textit{chain rule}, see Figure \ref{fig:chain}. It depicts the following integral
\begin{align}\label{chain}
	\int \mathcal{D} v\,
	\frac{1}{(z-\bar{v})^{\lambda}(v-\bar{w})^{\rho}} = a(\lambda, \rho) \,
	\frac{1}{(z-\bar{w})^{\lambda+\rho-2s}},
\end{align}
where the coefficient is given by
\begin{align}
	a(\lambda, \rho) = e^{-i\pi s}\,
	\frac{\Gamma(\lambda+\rho-2s)\Gamma(2s)}{\Gamma(\lambda)\Gamma(\rho)}.
\end{align}
The proof can be found in \cite[Appendix A]{DKM1}.

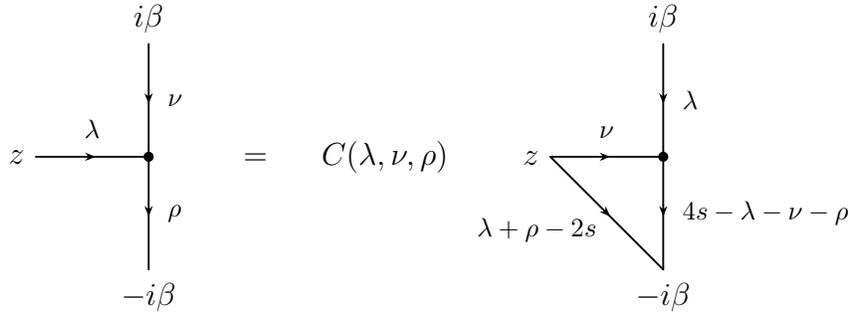
\begin{figure}[h]
	\centering
	\begin{tikzpicture}[thick, line cap = round]
	\def\l{1.5}
	\def\r{1.5pt}
	\draw[->-] (0,0) node[left] {$z$} to node[midway,above=3pt]{\footnotesize $\lambda$} (\l,0);
	\draw[fill = black] (\l,0) circle (\r);
	\draw[->-] (\l,0) to node[midway,right=3pt]{\footnotesize $\rho$} (\l,-\l) node[below = 1pt]{$-\imath \beta$};
	\draw[->-] (\l, \l) node[above]{$\imath \beta$} to node[midway, right=3pt]{\footnotesize $\nu$} (\l, 0);
	\end{tikzpicture}
	\begin{tikzpicture}[thick, line cap = round]
		\def\l{1.5}
		\def\r{1.5pt}
		\draw[->-] (0,0) node[left] {$\hspace{0.4cm} = \hspace{0.6cm} C(\lambda,\nu,\rho) \hspace{1cm} z$} to node[midway,above=3pt]{\footnotesize $\nu$} (\l,0);
		\draw[fill = black] (\l,0) circle (\r);
		\draw[->-] (\l,0) to node[midway,right=3pt]{\footnotesize $4s - \lambda - \nu - \rho$} (\l,-\l);
		\draw[->-] (\l, \l) node[above]{$\imath \beta$} to node[midway, right=3pt]{\footnotesize $\lambda$} (\l, 0);
		\draw[->-] (0,0) to node[pos = 0.65, left = 6pt]{\footnotesize $\lambda + \rho - 2s$} (\l, - \l) node[below = 1pt]{$-\imath \beta$};
	\end{tikzpicture}
	\caption{Euler transformation} \label{fig:Euler}
\end{figure}

The second identity is the \textit{Euler transformation}, see Figure \ref{fig:Euler}. It represents the following formula
\begin{align}\label{Euler}
	\begin{aligned}
		&\int \mathcal{D} w \;
		\frac{1}{(z-\bar{w})^{\lambda}(i\beta-\bar{w})^{\nu}(w+i\beta)^{\rho}} = C(\lambda,\nu,\rho)\; \\[8pt]
		&\quad \times (z+i\beta)^{2s-\lambda-\rho}\, \int \mathcal{D} w \; \frac{1}{(z-\bar{w})^{\nu}(i\beta-\bar{w})^{\lambda}(w+i\beta)^{4s-\lambda-\nu-\rho}},
	\end{aligned}
\end{align}
where we denoted
\begin{align}
C(\lambda,\nu,\rho) = (2i\beta)^{2s-\nu-\rho}\,\frac{\Gamma(4s-\lambda-\nu-\rho)\Gamma(\lambda+\nu+\rho-2s)}
{\Gamma(\rho)\Gamma(2s-\rho)}\,.
\end{align}
This identity is related to the Euler transformation of the hypergeometric function
\begin{equation}
	{}_2 F_1(a, b,c; x) = (1 - x)^{c - a - b} {}_2 F_1(c - a, c - b, c; x).
\end{equation}
Indeed, their equivalence easily follows from the following integral representation of the hypergeometric function
\begin{align}\label{F1}
	\begin{aligned}
		{}_2 F_1 \biggl(a, b, c ; \,  \frac{1}{2} + \frac{\imath z}{2\beta} \biggr) &=
		(2i\beta)^{a}\,e^{i\pi s}\,\frac{\Gamma(c)\Gamma(2s+a-c)}{\Gamma(a)\Gamma(2s)} \\[6pt]
		&\times\int \mathcal{D} w \,
		\frac{1}{(z-\bar{w})^b(i\beta-\bar{w})^{c-b}(w+i\beta)^{2s+a-c}}.
	\end{aligned}
\end{align}
This representation is depicted in Figure \ref{fig:2F1}, where we denoted
\begin{align}
	A(a,b,c) = (2i\beta)^{a}\,e^{i\pi s}\,\frac{\Gamma(c)\Gamma(2s+a-c)}{\Gamma(a)\Gamma(2s)}.
\end{align}

\begin{figure}[t]
	\centering
	\begin{tikzpicture}[thick, line cap = round]
		\def\l{1.5}
		\def\r{1.5pt}
		\draw[->-] (0,0) node[left] {$\displaystyle {}_2 F_1 \biggl(a, b, c ; \,  \frac{1}{2} + \frac{\imath z}{2\beta} \biggr) \; = \; A(a,b,c)\;\;\; z$} to node[midway,above=3pt]{\footnotesize $b$} (\l,0);
		\draw[fill = black] (\l,0) circle (\r);
		\draw[->-] (\l,0) to node[midway,right=3pt]{\footnotesize $2s + a - c$} (\l,-\l) node[below = 1pt]{$-\imath \beta$};
		\draw[->-] (\l, \l) node[above]{$\imath \beta$} to node[midway, right=3pt]{\footnotesize $c - b$} (\l, 0);
	\end{tikzpicture}
	\caption{The hypergeometric function} \label{fig:2F1}
\end{figure}

\noindent Let us derive the formula \eqref{F1}. Using the standard Euler representation for the hypergeometric function
\begin{align}\label{2F1-euler}
	{}_2 F_1 (a, b , c ;  x) =
	\frac{\Gamma(c)}{\Gamma(b)\Gamma(c-b)}\,
	\int_{0}^1 d t \; \frac{t^{b-1}(1-t)^{c-b-1}}{(1-tx)^{a}}
\end{align}
we can rewrite \eqref{F1} as the integral identity
\begin{align}
	\begin{aligned}
		&\int \mathcal{D} w \; \frac{1}{(z-\bar{w})^b(i\beta-\bar{w})^{c-b}(w+i\beta)^{2s+a-c}} \\[8pt]
		& \qquad = e^{-i\pi s}\,\frac{\Gamma(a)\Gamma(2s)}{\Gamma(b) \Gamma(c - b) \Gamma(2s+a-c)} \,  \int_0^1 dt \; \frac{t^{b-1}(1-t)^{c-b-1}}{ \bigl( t(z - \imath \beta) + 2\imath \beta \bigr)^{a}}.
	\end{aligned}
\end{align}
This identity is a special case of more general relation
\begin{align}\label{w-abc}
	\begin{aligned}
		&\int \mathcal{D} w \; \frac{1}{(z-\bar{w})^b(i\beta-\bar{w})^{c-b}(w-\bar{v})^{2s + a-c}} \\[8pt]
		& \quad = e^{-i\pi s}\,\frac{\Gamma(a)\Gamma(2s)}{\Gamma(b)\Gamma(c-b) \Gamma(2s + a - c)} \,  \int_0^1 dt \; \frac{t^{b-1}(1-t)^{c-b-1}}{ \bigl( t(z - \imath \beta) + \imath \beta  - \bar{v}\bigr)^{a}}
	\end{aligned}
\end{align}
when $v = \imath \beta$. The proof of \eqref{w-abc} is straightforward. At first step we use Feynman formula
\begin{align}
\frac{1}{A_1^{\alpha_1}\,A_2^{\alpha_2}} =
\frac{\Gamma(\alpha_1+\alpha_2)}{\Gamma(\alpha_1)\Gamma(\alpha_2)}\,
\int_{0}^1 d t\,\frac{t^{\alpha_1-1}(1-t)^{\alpha_2-1}}{(tA_1+(1-t)A_2)^{\alpha_1+\alpha_2}}
\end{align}
to joint two functions in denominator
\begin{align}
	\begin{aligned}
			&\int \mathcal{D} w \;
			\frac{1}{(z-\bar{w})^b(i\beta-\bar{w})^{c-b}(w - \bar{v})^{2s+a-c}}  \\[8pt]
			& \hspace{1cm}  = \frac{\Gamma(c)}{\Gamma(b)\Gamma(c-b)} \int_{0}^1 d t \;  t^{b-1} (1-t)^{c-b-1} \\[6pt]
			& \hspace{1cm} \times	\int \mathcal{D} w \frac{1}{(tz+(1-t)i\beta - \bar{w})^{c}(w - \bar{v})^{2s+a-c}}.
	\end{aligned}
\end{align}
Secondly, we apply the chain rule \eqref{chain} for the integral in the last line
\begin{align}
	\begin{aligned}
		&\int \mathcal{D} w \; \frac{1}{(tz+(1-t)i\beta - \bar{w})^{c}(w - \bar{v})^{2s+a-c}} \\[8pt]
		& \quad = e^{-i\pi s}\,
		\frac{\Gamma(a)\Gamma(2s)}{\Gamma(c)\Gamma(2s+a-c)} \; \frac{1}{(tz+(1-t)i\beta  - \bar{v})^{a}} .
	\end{aligned}
\end{align}
and arrive at the formula \eqref{w-abc}.

\section{Frassek--Giardin\`a--Kurchan formula}\label{App:FGK}

In this appendix we clarify the relation between the formula for the reflection operator derived in Section \ref{sec:refl}
\begin{align}\label{Kf11}
	\mathcal{K}(s,x) = \frac{\Gamma \bigl( N + \frac{x - s + 1}{2} + \frac{\alpha}{\beta} 	\bigr)}{\Gamma \bigl( N + \frac{s - x + 1}{2} + \frac{\alpha}{\beta} \bigr)}, \quad
	N = \frac{1}{2\imath \beta} \bigl[ (z^2 + \beta^2) \partial_z + (s + x)z \bigr]
\end{align}
with the one obtained in \cite[Section 3.3]{FGK}. First, we note that the $K$-matrix considered in the work \cite{FGK}
\begin{align}
	\widetilde{K} \biggl( u + \frac{1}{2} \biggr) =
	\begin{pmatrix}
		q_1 + q_2 u  & q_3 u \\[6pt]
		q_4 u & q_1 - q_2 u
	\end{pmatrix}
\end{align}
contains two more parameters than ours  \eqref{K}
\begin{align}
	K\biggl(u + \frac{1}{2}\biggr) =
	\begin{pmatrix}
		\imath \alpha & u \\[6pt]
		-\beta^2 u & \imath \alpha
	\end{pmatrix}.
\end{align}
However, these additional parameters can be removed by conjugation
\begin{align}
	\widetilde{K} \biggl( u + \frac{1}{2} \biggr) =
	 G
	\begin{pmatrix}
		q_1 & u\\[6pt]
		(q_2^2 + q_3 q_4) \, u & q_1
	\end{pmatrix}
	G^{-1}
\end{align}
with the matrix
\begin{equation}
	G = e^{- (q_2/q_3)\sigma_-} \; q_3^{\sigma_3/2}, \qquad
	\sigma_- =
	\begin{pmatrix}
		0 & 0 \\
		1 & 0
	\end{pmatrix}, \qquad \sigma_3 =
	\begin{pmatrix}
		1 & 0\\
		0 & -1
	\end{pmatrix}.
\end{equation}
For the Lax matrix \eqref{L}
\begin{align}
	L(u) = \begin{pmatrix}
		u + S & S_- \\
		S_+ & u - S
	\end{pmatrix} =
	\begin{pmatrix}
		u + z\partial_z + s & - \partial_z \\
		z^2 \partial_z + 2sz & u - z \partial_z - s
	\end{pmatrix}
\end{align}
this conjugation is equivalent to the transformation of coordinate $z$
\begin{align}
	G^{-1} \, L(u) \,G = e^{(q_2/q_3) \partial_z} \; q_3^{z \partial_z} \;\, L(u) \,\; q_3^{-z\partial_z} \; e^{-(q_2/q_3) \partial_z},
\end{align}
where
\begin{align}
	e^{(q_2/q_3) \partial_z} \, \psi(z) = \psi(z - q_2/q_3), \qquad q_3^{z\partial_z} \, \psi(z) = \psi(q_3 z).
\end{align}
Therefore if we identify the parameters in $K$-matrices
\begin{equation}
	q_1 = \imath \alpha, \qquad q_2^2 + q_3 q_4 = - \beta^2,
\end{equation}
the corresponding monodromy matrices \eqref{Tn}
\begin{align}
	& T(u) = L_n(u) \cdots L_1(u) K(u) L_1(u) \cdots L_n(u) =\begin{pmatrix}
		A(u) & B(u) \\
		C(u) & D(u)
	\end{pmatrix}, \\[6pt]
	& \widetilde{T}(u) = L_n(u) \cdots L_1(u)  \widetilde{K} (u) L_1(u) \cdots L_n(u) =\begin{pmatrix}
		\widetilde{A}(u) & \widetilde{B}(u) \\
		\widetilde{C}(u) & \widetilde{D}(u)
	\end{pmatrix}
\end{align}
are related by similarity transformation
\begin{align}\label{TtT}
	\widetilde{T}(u) =  Z \, G \, T(u) \, G^{-1} \, Z^{-1} ,
\end{align}
where the operator $Z$ shifts and rescales all coordinates
\begin{align}
	Z = e^{ (q_2/q_3)( \partial_{z_1} + \, \dots \, + \partial_{z_n}) } \; q_3^{z_1 \partial_{z_1} + \, \dots \, + z_n \partial_{z_n}}.
\end{align}
Note that from the relation \eqref{TtT} we, in particular, obtain
\begin{align}
	q_3 \, \widetilde{B}(u) =  Z \, B(u) \, Z^{-1},
\end{align}
so that the operators $\widetilde{B}(u)$ and $B(u)$ are equivalent modulo change of coordinates.

Next note that the defining equation for our reflection operator \eqref{Kdef} can be rewritten in the form
\begin{align}\label{KMKM}
	\begin{aligned}
		&\mathcal{K}(s,x) \; M( u + v) \, K(u) \, M( u - v ) \\[6pt]
		& = M( u - v) \, K(u) \, M( u + v ) \; \mathcal{K}(s,x),
	\end{aligned}
\end{align}
where $M(u)$ is the Lax matrix with the generators of spin $(s + x)/2$
\begin{align}
	M(u) = \begin{pmatrix}
		u + J & J_- \\[6pt]
		J_+ & u - J
	\end{pmatrix} =
	\begin{pmatrix}
		u + z\partial_z + \frac{s+x}{2} & - \partial_z \\[6pt]
		z^2 \partial_z + (s+x)z & u - z \partial_z - \frac{s+x}{2}
	\end{pmatrix}
\end{align}
and $v = (x - s)/2$. The equation \eqref{KMKM} is equivalent to the one considered in \cite[eq. (3.13)]{FGK} due to the Lax matrix property
\begin{align}
	M(u) \, M(1-u) = (u - s)(1-u-s) \bm{1}.
\end{align}

Finally, we can rewrite our reflection operator in the spirit of the work \cite[eq. (3.23), (3.25)]{FGK}
\begin{align} \label{K-Fr}
			\mathcal{K}(s,x)=  e^{- \frac{1}{\imath \beta} \, J_+} \,  e^{\frac{\imath \beta}{2 } \, J_-} \, \frac{\Gamma\bigl( J + \frac{x - s}{2} + \frac{1}{2} + \frac{\alpha}{\beta} \bigr)}{ \Gamma\bigl( J - \frac{x - s}{2} + \frac{1}{2} + \frac{\alpha}{\beta} \bigr) } \, e^{- \frac{\imath \beta}{2 } \, J_-} \, e^{ \frac{1}{\imath \beta} \, J_+}.
\end{align}
The equivalence to the formula \eqref{Kf11} follows from the identity
\begin{align}\label{N2}
	N = \frac{1}{2\imath \beta}(J_+ - \beta^2 J_-) = e^{- \frac{1}{\imath \beta} \, J_+} \,  e^{\frac{\imath \beta}{2 } \, J_-} \, J \, e^{- \frac{\imath \beta}{2 } \, J_-} \, e^{ \frac{1}{\imath \beta} \, J_+}.
\end{align}
This identity can be directly checked using relations
\begin{align}
	e^{\lambda J_\pm} \, J \, e^{-\lambda J_\pm} = J \mp \lambda J_\pm, \qquad e^{\lambda J_+} \, J_- \, e^{- \lambda J_+} = J_- + 2\lambda J - \lambda^2 J_+,
\end{align}
which are easily proved differentiating left-hand sides with respect to $\lambda$ and using commutation relations for $J_a$.

\section{Hypergeometric functions relation} \label{App:Ind}

Let us prove the formula \eqref{Psi-Phi} relating hypergeometric functions with different arguments
\begin{align}\label{Psi-Phi2}
	\begin{aligned}
		&{}_2 F_1 \biggl( s+ \imath \lambda, s - \imath \lambda, s + g; \, \frac{1}{2} + \frac{\imath z}{2\beta} \biggr) \\[6pt]
		& \qquad = \frac{ (2\beta)^{2s} \, \Gamma(2s)}{\Gamma(s + \imath \lambda) \, \Gamma(s - \imath \lambda)} \, \int_0^\infty  dy \,\, y^{s + g - 1} \, (1 + y)^{s - g} \\[6pt]
		&\qquad  \times \frac{e^{\imath \pi s}}{\bigl( z + \imath \beta(1 + 2y) \bigr)^{2s}} \; {}_2 F_1 ( s+ \imath \lambda, s - \imath \lambda, s + g; \, -y )
	\end{aligned}
\end{align}
where $\Im z > 0$. We start from the identity \cite[eq. (2.1)]{N}
\begin{align}\label{Ner}
	\begin{aligned}
		\int_0^\infty &dy \; \frac{y^{\alpha - 1}}{(y + z)^\rho} \, {}_2 F_1(p,q,r; - y) = \frac{\Gamma(r)}{\Gamma(p) \, \Gamma(q) \, \Gamma(\rho)} \; z^{\alpha - \rho}  \\[10pt]
		& \times  \int_{\imath \mathbb{R}} \frac{dt}{2\pi\imath } \; \frac{\Gamma(t + \alpha) \, \Gamma(\rho - t - \alpha) \, \Gamma(p + t) \, \Gamma(q + t) \, \Gamma(-t)}{\Gamma(r + t)} \, z^t
	\end{aligned}
\end{align}
with five parameters $\alpha, \rho, p, q ,r$. As it is explained in \cite{N}, this identity can be easily proved using Parseval's formula for Mellin transform. In the case $\alpha = r$ integral from the right represents the hypergeometric function \cite[\href{https://dlmf.nist.gov/15.6.7}{(15.6.7)}]{DLMF}
\begin{align}
	\begin{aligned}
		&\int_{\imath \mathbb{R}} \frac{dt}{2\pi\imath } \; \Gamma(\rho - t - \alpha) \, \Gamma(p + t) \, \Gamma(q + t) \, \Gamma(-t) \, z^t \\[6pt]
		&\; = \frac{\Gamma(p) \, \Gamma(q) \, \Gamma(\rho - r + p) \, \Gamma(\rho - r + q)}{\Gamma(\rho - r + p + q)} \, {}_2 F_1(p, q, \rho - r + p + q; 1 - z).
	\end{aligned}
\end{align}
Inserting this representation into \eqref{Ner} and performing Euler transformation
\begin{align}
	{}_2 F_1(a,b,c;z) = (1 - z)^{c - a - b}	{}_2 F_1(c-a,c-b,c;z)
\end{align}
from both sides we arrive at
\begin{align}
	\begin{aligned}
		&\int_0^\infty dy \; \frac{y^{r - 1}}{(y + z)^{r - p - q}} \; (1 + y)^{r - p - q} \; {}_2 F_1(r - p, r - q, r; -y)  \\[10pt]
		& \hspace{1.5cm} = \frac{\Gamma(r) \, \Gamma(\rho - r + p) \, \Gamma(\rho - r + q)}{\Gamma(\rho) \, \Gamma(\rho - r + p + q)}\\[6pt]
		& \hspace{1.5cm}  \times{}_2 F_1(\rho - r + p, \rho - r + q, \rho - r + p + q; 1 - z).
	\end{aligned}
\end{align}
This identity is equivalent to \eqref{Psi-Phi2} under identification
\begin{align}
	r = s + g, \qquad \rho = 2s, \qquad p = - \imath \lambda + g, \qquad q = \imath \lambda + g
\end{align}
and change of variable
\begin{equation}
	z \to \frac{1}{2} - \frac{\imath z}{2\beta}.
\end{equation}

\end{document}